\documentclass[]{article}
\usepackage{emulateapj}
\usepackage{epsfig}
\usepackage{ifthen}
\submitted{Draft}

\newcommand{\HI}{\ion{H}{1}} 
\newcommand{\FeII}{\ion{Fe}{2}} 
\newcommand{\CIII}{\ion{C}{3}} 
\newcommand{\CIV}{\ion{C}{4}} 
\newcommand{\SiIII}{\ion{Si}{3}} 
\newcommand{\SiIV}{\ion{Si}{4}} 
\newcommand{\NV}{\ion{N}{5}} 
\newcommand{\OVI}{\ion{O}{6}} 

\newcommand{\msun}{{\rm M}_\odot}

\def\gsim{\;\rlap{\lower 2.5pt
 \hbox{$\sim$}}\raise 1.5pt\hbox{$>$}\;}
\def\lsim{\;\rlap{\lower 2.5pt
   \hbox{$\sim$}}\raise 1.5pt\hbox{$<$}\;}

\def\kms{\rm\,km\,s^{-1}}
\def\mpc{{\rm\,Mpc}}

\def\cm{{\rm\,cm}}

\def\lya{{{\rm Ly}\alpha}}

\def\kms{{\rm\,km\,s^{-1}}}
\def\K{{\rm\,K}}
\def\spose#1{\hbox to 0pt{#1\hss}}
\def\lta{\mathrel{\spose{\lower 3pt\hbox{$\mathchar''218$}}
     \raise 2.0pt\hbox{$\mathchar''13C$}}}
\def\gta{\mathrel{\spose{\lower 3pt\hbox{$\mathchar''218$}}
     \raise 2.0pt\hbox{$\mathchar''13E$}}}

\def\zsol{{\,Z_\odot}}

\begin{document}
	
\title{Metallicity of the intergalactic medium using pixel
statistics:\\ I. Method} 

\author{Anthony Aguirre and Joop Schaye} \affil{School of Natural
Sciences, Institute for Advanced Study, Princeton, NJ 08540\\
aguirre@ias.edu, schaye@ias.edu} \vskip0.1in \author{Tom Theuns}
\affil{Institute of Astronomy, Madingley Rd., Cambridge CB3 0HA, UK\\
tt@ast.cam.ac.uk}

\setcounter{footnote}{0}

\begin{abstract}

Studies of absorption spectra of high-$z$ QSOs have revealed that the
intergalactic medium at $z\sim 2-3$ is enriched to $\sim
10^{-3}-10^{-2}\zsol$ for gas densities more than a few times the mean
cosmic density, but have not yet produced an accurate metallicity
estimate, nor constrained variations in the metallicity with density,
redshift, or spatial location.  This paper discusses the ``pixel
optical depth'' (POD) method of QSO spectrum analysis, using realistic
simulated spectra from cosmological simulations.  In this method,
absorption in $\lya$ is compared to corresponding metal absorption on
a pixel-by-pixel basis, yielding for each analyzed spectrum a single
statistical correlation encoding metal enrichment information.  Our
simulations allow testing and optimization of each step of the
technique's implementation.  Tests show that previous studies have
probably been limited by \CIV\ self-contamination and \OVI\
contamination by \HI\ lines; we have developed and tested an effective
method of correcting for both contaminants.  We summarize these and
other findings, and provide a useful recipe for the POD technique's
application to observed spectra.  Our tests reveal that the POD
technique applied to spectra of presently available quality is
effective in recovering useful metallicity information even in
underdense gas. We present an extension of the POD technique to
directly recover the intergalactic metallicity as a function of gas
density. For a given ionizing background, both the oxygen and carbon
abundance can be measured with errors of at most a factor of a few
over at least an order of magnitude in density, using a single
high-quality spectrum.

\end{abstract}
\keywords{cosmology: theory --- intergalactic medium ---
quasars: absorption lines}

\section{Introduction}
\label{sec-intro}

	Studies of absorption lines in QSO spectra have established
that at high redshift ($z \ga 2$) the intergalactic medium (IGM) is
enriched by metals to metallicity $10^{-3} \la Z/\zsol \la 10^{-2}$
at densities above a few times the cosmic mean (e.g., Tytler
et al.\ 1995; Cowie et al.\ 1995; Songaila \& Cowie 1996; Rauch,
Haehnelt, \& Steinmetz 1997; Dav\'e et al.\ 1998).  Since stars are
thought to form in appreciable quantities only inside much denser mass
condensations, the observed enrichment indicates that galaxies, which
form from the IGM, also feed back some of their nucleosynthetic
products in a process which is currently poorly understood but which
is likely to be a crucial ingredient of galaxy formation and
evolution.

	Because galaxies form in relatively high-density regions,
metals in low-density regions must have traveled a significant
distance.  This puts strong constraints on the intergalactic (IG)
enrichment mechanism, therefore determining the IG metallicity at the
lowest possible densities is crucial for understanding the feedback of
metals from galaxies to the IGM (see, e.g., Aguirre et al.\ 2001;
Madau, Ferrara \& Rees 2001).

	Previous studies, performed primarily with the Keck HIRES
instrument, have used measured metal line column densities to show
that, at $z\approx 3$, most absorbers with $\log N({\rm H\,I}) > 14.5$
(corresponding to gas overdensities $\delta \equiv
\rho/\langle\rho\rangle \ga 5-10$ [Schaye 2001]) have associated
carbon lines (e.g., Tytler at al. 1995; Cowie et al.\ 1995; Songaila
\& Cowie 1996; Ellison et al. 2000).  These studies have not, however,
made clear how this metallicity changes with spatial location,
redshift, or gas density.  Studies employing line-fitting could in
principle treat the first two issues.  But pushing to lower gas
densities where individual metal lines are undetectable requires a
statistical analysis; at high redshift this is also required for
metals absorption falling blue-wards of $\lya$, where the metals lines
are difficult to disentangle from \HI\ absorption.

Ellison et al.\ (1999, 2000; see also Cowie \& Songaila 1998) have
compared two ways of doing this: first by stacking the spectra of the
metal-line regions corresponding to a set of low column density $\lya$
lines, or second by comparing the $\lya$ optical depth in each pixel with
the optical depth in the pixel at the corresponding metal-line
wavelength.  The latter ``pixel optical depth'' (POD) method, which
was devised and first used by Cowie \& Songaila (1998), has
several advantages: it is fast, it is objective, it preserves more of
the spectrum's information, it can be applied to heavily contaminated
regions, and it appears to be more sensitive to metals in low-density
gas, even in uncontaminated regions.  Its shortcoming, however, has
been that it is not immediately clear how to interpret the results of
the POD method, and it is unclear how to convert the set of PODs into
a metallicity $Z$ at a given overdensity $\delta$, especially when
contamination is important, as for \OVI\ (see Schaye et al.\ 2000a).

The purpose of the present study is to present several important
improvements to and extensions of the POD technique, to provide a
complete and useful explanation of the method, and to test it using
realistic spectra generated from hydrodynamic cosmological
simulations. The method as described here will be applied to a sample
of observed QSO spectra in a future paper. The present paper should
also serve as a useful reference for prospective users of the POD
technique.  The testing we perform allows us to estimate the accuracy
with which \HI\ and metal line PODs can be recovered (and to refine
the method for doing so), and to draw the connection between the
optical depths and $Z(\delta)$ for the gas. By comparing recovered
quantities directly to the true (simulated) ones, we demonstrate that
the POD method does work, and that we can interpret directly the
meaning of features in the POD statistic.

Section~\ref{sect-overview} contains the basics of the method,
pointing forward to figures in the main text, for illustration. This
section is meant for readers not interested in all the details, or as
a general overview for readers unfamiliar with pixel statistics.  The
full details of the method, and of the tests we performed, can be
found in the following sections.  In \S~\ref{sec-spgen} we describe
the simulations used in this study and how we generate spectra from
them. In \S~\ref{sec-podmeth} we give a careful account of the
recovery of metal and \HI\ PODs from QSO spectra, testing the
individual steps by comparing recovered and ``true'' optical depths;
this section can be used as a reference for the POD method of spectrum
analysis, and is summarized in the Appendix. Section~\ref{sec-interp}
discusses and tests the interpretation of the recovered PODs:
Sections~\ref{sec-hiinterp} and~\ref{sec-zinterp} relate the \HI\ and
metal PODs to the density of the absorbing gas, and
\S~\ref{sec-restest} shows the relation between metal PODs and the
assumed gas metallicity, given various levels of noise and other
uncertainties.  In \S~\ref{sec-invert} we formulate and test a
procedure whereby $Z(\delta)$ can be recovered directly and perform
simple tests of this procedure. Finally, we discuss our results and
conclude in \S~\ref{sec-conc}.

\section{Overview of the method}
\label{sect-overview}
The aim of the method is to {\em statistically} obtain the physical
properties of the gas that produces the absorption in a given redshift
range. The first step is to infer, in the presence of noise,
contamination, saturation, etc., for each spectrum pixel the best
possible estimate of the optical depth $\tau_{\lya}$ due to $\lya$
aborption by gas at some redshift, as well as the optical depths
$\tau_Z$ for various metal transitions in gas at that same redshift.
The second step converts the inferred optical depths to physical gas
densities, using the tight relation that exists between them. This
yields density ratios of \HI\ to various metal species such as \CIV\
and \OVI.  Ultimately, however, we aim to measure the metallicity as a
function of overdensity. So thirdly, we use ionization corrections (as
determined by the photoionization package CLOUDY), for gas at given
density and temperature illuminated by a chosen ionizing background.

{\em Simulated spectra.} We use simulated spectra generated from
hydrodynamical simulations to test and illustrate the method. We patch
together the physical state of the gas along many uncorrelated
sight-lines through the simulation box, in order to obtain one long
sight-line of redshift extent $z\sim 1.5-z_{\rm qso}$, where $z_{\rm
qso}$ is the assumed emission redshift of the quasar. (We illustrate
the method for $z_{\rm qso}=2.5$ and 3.5). Metals are distributed by
hand, for example by assuming a uniform metallicity of 1\% solar. We
then generate simulated spectra (using our chosen ionizing background
to compute ionization balances), taking into account all hydrogen
Lyman transitions and absorption by important metal transitions. The
wavelength extent and signal-to-noise ratio of these simulated spectra
are chosen to be similar to the very high quality echelle spectra for
bright quasars, as obtained by HIRES on Keck, or UVES on the VLT.
Example spectra are shown in
Figures~\ref{fig-sampspec}-\ref{fig-sampspec_z2.5}.

{\em Recovering the optical depth.} The primary issue in recovering
the optical depth $\tau$ from the flux $F={\rm exp}(-\tau)$ is that
information is lost when the optical depth is either too high, too
low, or contaminated by other transitions. For hydrogen, we use the
higher order transitions to recover the optical depth in saturated
pixels. However, these higher order pixels are (potentially)
contaminated by other hydrogen transitions from lower redshifts. (Note
that our simulated spectra have a realistic level of this type of
contamination, because they are generated from realistically long
sight-lines through simulations that reproduce the observed evolution
of the Ly$\alpha$ forest.) For hydrogen, we therefore consider only
those pixels of the Lyman series which are neither too black, nor too
close to the continuum. The optical depth of that redshift pixel is
then the minimum optical depth of all the transitions that qualify,
thereby minimizing contamination both by Lyman series lines from lower
redshift, by from other transitions. Figure~\ref{fig-rectest} shows
how well the recovered optical depth matches the true one.

For metal transitions occuring in the $\lya$ forest, correcting for
contamination is of course crucial. For \OVI\ and \NV, we use the
$\lya$ optical depth determined earlier to correct for contamination
by higher-order \HI\ lines. We further reduce contamination by using
the minimum optical depth for each of the components of the
doublet. For \CIV\ and \SiIV, it is possible -- and very important --
to correct for self-contamination (i.e., where the second component of
the doublet happens to fall on top of the first component of another
system), and we find that an iterative correction works well (see
Figure~\ref{fig-recztest_civ}). On a pixel-by-pixel basis the recovery is
good but imperfect, especially at low optical depth; recovered \OVI\
and \NV\ PODS are also systematically high, due to contamination (see
Fig.~\ref{fig-recztest_ovi}).  To obtain a statistical signal we bin
the recovered metal PODs according to the corresponsing hydrogen
optical depth, and find the median metal POD in each bin, as
illustrated, for example, in Figures~\ref{fig-odtest_ovi} and
\ref{fig-odtest_civ}.

\begin{figure*}
\vbox{ \centerline{ \epsfig{file=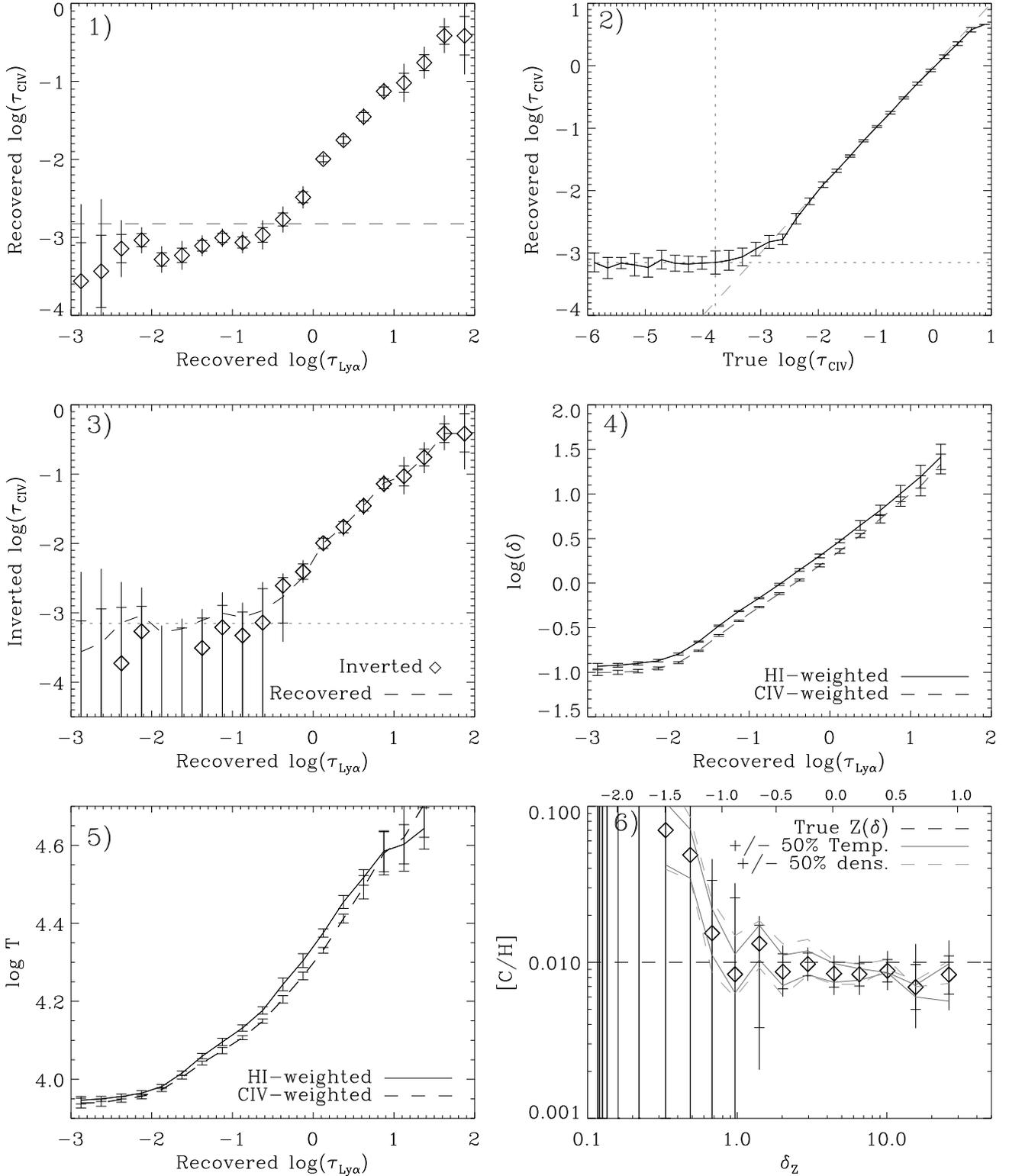,width=18.0truecm}}
\vskip0.1in
\figcaption[]{ \footnotesize Test of the recovery of gas metallicity
versus density.  Here, $z_{\rm qso}=2.5$ and we recover carbon. The
analyzed spectrum has $Z=0.01\zsol$, whereas the spectra used in
performing the inversion use $Z=0.001\delta\zsol$.  {\bf Panel 1:}
$\tau_{\rm CIV}^{\rm rec}$ versus $\tau_{\lya}^{\rm rec}$ (diamonds).
1- and  2-$\sigma$ errors determined using bootstrap resampling are
shown.  {\bf Panel 2:} 
The median binned `true' vs. `recovered' metal POD relation.  Errors
are computed by taking the standard deviations among 10 independent
spectra. 
The horizontal (vertical) light dotted line indicates the recovered
(true) $\tau_{\rm CIV}$ value
below which the relation is taken to be flat. {\bf Panel 3:}
`Inverted' metal PODs using panels 1 and 2, as described in
text. Dotted line is as in panel 2.  The dashed line re-plots the
relation of panel 1 for comparison.  {\bf Panel 4:} $\lya$ (solid
line) and metal (dashed line) optical depth weighted overdensity of
absorbing gas vs.\ recovered $\lya$ POD.  Errors are computed as in
panel 2. {\bf Panel 5:} As panel 3 but for temperature.  {\bf Panel 6:} Recovered
carbon abundance in solar units versus metal optical depth weighted
density, given an ionization correction using density and temperature
from panels 3 and 4, and the same ionizing background as was used in the
generation of the spectra.  The true metallicity is plotted
(dashed line) for comparison.  Solid and light dashed lines indicate,
respectively, changes induced by an overall $\pm 50\%$ scaling in
temperature or density in the ionization correction.
\label{fig-inverttest}}}
\vspace*{0.5cm}
\end{figure*}

These binned, recovered PODs can be further improved using the
simulations. By employing simulated spectra with the same properties
as the given spectrum to compute the relation between ``true'' and
``recovered'' PODs, we can ``invert'' the recovered PODs (which are
affected by contamination, noise, etc.)  to obtain an estimate of the
``true'' PODs in the given spectrum.  This process is illustrated in
Fig.~\ref{fig-inverttest}.

{\em From optical depth to density.} In the simulations (and
presumably in reality), there is a tight relation between optical
depth and density, both for hydrogen (see figure~\ref{fig-tauvdel})
and for metals. These relations allow us to invert the optical depths
and obtain the underlying (redshift space) densities.

{\em Ionization corrections.} To convert absorption densities to
physical densities, we require ionization corrections, which depend on
the shape and amplitude of the ionizing background, as well as on the
density and temperature. Fortunately, the latter two quantities 
are related in the simulations, and our simulations produce a
temperature-density relation that is consistent with the relation
measured by Schaye et al.~(2000b) from the
widths of $\lya$ lines as a function of their column density.
Therefore, assuming an ionizing background (normalized to match
the mean observed absorption by the IGM), we can finally deduce true
physical densities of various atomic species, obtaining for example
the oxygen abundance as a function of the hydrogen (over) density.

By applying this method to simulated spectra, we can estimate the
importance of various physical effects and modeling uncertainties.
For example, we investigate the effects of uncertainties in the
temperature-density and density-column density relations, continuum
fitting errors, partial wavelength coverage and noise. We demonstrate
that despite these uncertainties, the algorithm is very powerful.  In
Figs.~\ref{fig-inverttest} and~\ref{fig-inverttest_multi} we further show
that we can accurately recover the metallicity of gas down to (at
least) the cosmic mean gas density, given an accurate estimate of the
ionizing bakground.

We conclude that relative to other techniques of analyzing absorption
spectra, there are several advantages to the POD method. First,
because it is not based on line fitting, the analysis is extremely
fast, objective, and its implementation is straightforward. Second, it
is accurate and robust because it uses the information of all
transitions for each redshift pixel. Third, absorption can be detected
{\em statistically}, enabling us to probe lower optical depths, which
is crucial for studying low levels of metal contamination. We now
proceed to describe and test the various steps in more detail.

\section{Generation of spectra from simulations}
\label{sec-spgen}
To test the optical depth technique we require
realistic absorption spectra with wavelength range,
resolution, pixel size, and noise properties similar to the observed
spectra. Besides absorption by $\lya$ and the metal transitions of
interest, the spectra must contain absorption from contaminants
such as higher-order Lyman lines and additional metal lines. Before
describing the procedure used to produce such spectra from a
cosmological simulation, we will briefly summarize the properties of
the simulation itself.

In this study we employ a hydrodynamical simulation of the currently
popular flat, scale invariant, vacuum energy dominated cold dark
matter model. The simulation was performed with a modified version of
HYDRA (Couchman, Thomas, \& Pearce 1995) and uses smooth particle
hydrodynamics (SPH). The evolution of a periodic, cubic region of the
universe of comoving size $12~h^{-1}~\mpc$ was followed to redshift
$z=1.5$. The model has a total matter density $\Omega_m = 0.3$, vacuum
energy density $\Omega_\Lambda = 0.7$, baryon density $\Omega_b h^2 =
0.019$, Hubble constant $H_0 = 65~\kms\,\mpc^{-1}$, and the amplitude
of the initial power spectrum is normalized to $\sigma_8 = 0.9$. The
simulation uses $256^3$ gas particles and $256^3$ cold dark matter
particles, yielding particle masses of $2.0\times 10^6~\msun$ and
$1.1\times 10^7~\msun$ respectively. This mass resolution has been
demonstrated to be sufficient to resolve the $\lya$ forest spectra
(Theuns et al.~1998; Bryan et al.~1999; Schaye et al.~2000b). Gas
particles are converted to collisionless star particles if they
satisfy the following three conditions: (1) the density exceeds 80
times the mean baryon density; (2) the density $\rho/m_H$ exceeds
$10^{-2}~\cm^{-3}$; (3) the temperature is less than $2\times
10^4~\K$. Feedback from star formation is not included. The IGM is
assumed to be of primordial composition with a helium abundance of
0.24 by mass (metals are added only during the generation of spectra
from the completed simulation and do therefore not contribute to the
cooling of the gas). The gas is photoionized and photoheated by a
model of the UV-background, designed to match the evolution of the
mean $\lya$ absorption and the temperature-density relation measured
by Schaye et al.~(2000b).

The final spectra have characteristics typical of spectra taken with
HIRES/KECK or UVES/VLT.  The spectra span a wavelength range
3000-7500~\AA\ and include absorption by gas from redshift $z=1.5$ to
the redshift of the QSO ($z_{\rm qso}$), which is set to either 2.5 or
3.5. Absorption from Ly1 ($\lya$,$\lambda$1216) to Ly31, \CIII\
($\lambda$977), \CIV\ ($\lambda\lambda$1548, 1551), \NV\
($\lambda\lambda$1239, 1243), \OVI\ ($\lambda\lambda$1032, 1038),
\SiIII\ ($\lambda$1207), \SiIV\ ($\lambda$1394, 1403), \FeII\
($\lambda$ 1145, 1608, 1063, 1097, 1261, 1122, 1082, 1143, 1125) is
included. These transitions account for nearly all absorption by
$z=1.5$-3.5 absorbers in the 3000-7500~\AA\ window. Other commonly
observed absorption lines, mainly from neutral and singly ionized
species, occur in gas of densities higher than the maximum gas density
in the simulation. However, with the
exception of damped $\lya$ absorption, the observed spectral filling
factor of this gas is small enough that the effect of these absorbers
is negligible for our purposes. The simulated spectra include
absorption from absorbers with $z>1.5$ only.  However, contamination by
absorbers with redshift $z<1.5$ is minimal because nearly all commonly
observed absorption lines have rest wavelengths in the far UV.

At many hundreds of snapshot times, the position, density, temperature
and velocity of those gas particles that affect the physical state of
the gas along 6 randomly chosen sight-lines through the
simulation box (i.e., those particles whose SPH smoothing kernel intersects
the sight-line) are saved. The time between outputs is similar to the
time it takes light to cross the simulation box.  A long, continuous
sight-line spanning $z=1.5$ to $z_{\rm qso}$ is created by connecting
short sight-lines, choosing at random from those sight-lines with
redshifts close to the desired redshift. Before connecting a short
sight-line it is cycled periodically so that the point at which the
inferred $\lya$ transmission is maximum is at the ends. Since the
simulation box is sufficiently large for each short sight-line to have
a maximum in the $\lya$ transmission close to 1 (i.e., no absorption),
there are generally no strong discontinuities in the long sight
line. After a short sight-line has been inserted in the continuous
sight-line, the gas density along the sight-line is scaled to account
for the Hubble expansion during the time it takes light to cross the
simulation box (this scaling is not very important because the
simulation box is small). Correlations in physical quantities on scales
greater than the box size are of course not modeled correctly. However,
for the purpose of studying metal absorption all that matters is that
the contamination by absorbers at various redshifts is modeled
realistically.

The ionization balance of each gas particle is computed using the
publicly available photoionization package
CLOUDY\footnote{\texttt{http://www.pa.uky.edu/$\sim$gary/cloudy/}}
(version 94; see Ferland 2000 for details), assuming the gas to be
optically thin. The gas is illuminated by the redshift-dependent model
of the UV/X-ray background of Haardt \& Madau (1996), updated in
Haardt \& Madau (2001)\footnote{The data and a description of the
input parameters can be found at
\texttt{http://pitto.mib.infn.it/$\sim$haardt/refmodel.html}}, which
includes contributions from QSOs and galaxies. The amplitude of the
background radiation is scaled so that the mean $\lya$ absorption
matches the measurements of Schaye et al.~(2000b)\footnote{Note that
the standard practice of rescaling the spectrum (i.e., multiplying the
optical depth by a fixed factor) instead of the amplitude of the
background radiation, could lead to large errors if absorption by
metals is included.}. The UV/X-ray background used to compute the
spectra is different from the model of the background radiation that
was used in the hydrodynamical simulation. The exact UV/X-ray
background used in the simulation is, however, unimportant as long as
the resulting thermal evolution of the IGM roughly matches the
observations, which is indeed the case. The gas fraction in each
ionization state (e.g., \CIV/C) depends on the density, temperature,
and redshift.

For each short sight-line the absorption spectra of the various
transitions, i.e., the optical depths as a function of redshift,
$\tau_i(z)$, are computed using the formalism described in Theuns et
al.~(1998). For each transition $i$ with rest wavelength $\lambda_i$,
the spectrum $\tau_i(z)$ is then added to the final, continuous
spectrum $\tau(\lambda)$ using $\lambda = \lambda_i(1+z)$.  The final,
continuous spectrum $\tau(\lambda)$ is converted into a normalized,
flux spectrum using $F = \exp(-\tau)$. It is then processed in three
steps to give it characteristics similar to observed spectra taken
with HIRES/KECK and UVES/VLT. First, it is convolved with a Gaussian
with FWHM of $6.6~\kms$ to mimic instrumental broadening. Second, it
is resampled onto 0.04~\AA\ pixels. Third, noise is added.  For
simplicity we have assumed the noise to be Gaussian and to consist of
two components, one with flux-independent rms amplitude
$(1/3)(S/N)^{-1}$ and one with amplitude $(2/3)F(\lambda)(S/N)^{-1}$,
where $S/N$ is the signal-to-noise ratio in the continuum. For most
tests we quote results with both $S/N=25$ and $S/N=100$. Of course,
when comparing with real data one should impose noise that is
statistically equivalent, both in its dependence on wavelength and
flux, to the noise in the observations.

To facilitate a physical interpretation of the results presented in
this work, we have also computed the gas density along the continuous
sight-line.  Because of redshift space distortions, multiple gas
elements can contribute to the optical depth of a given pixel. We
therefore follow Schaye et al.~(1999) and use optical depth weighted
physical quantities. The density of a pixel in velocity space is
defined as the sum, weighted by the contribution to its optical depth,
of the density of all the gas elements that contribute to the
absorption in that pixel by a given transition. Note that because
different elements have different ionization corrections, the density
weighted by, say, the \CIV\ optical depth will generally differ
somewhat from the density weighted by the \HI\ optical depth, even if
the metallicity is uniform. This is not an artifact of the density
definition; there is no reason why the \CIV\ and \HI\ absorption at a
given redshift should arise in exactly the same gas.

\section{The optical depth technique}
\label{sec-podmeth}
	We now describe the POD analysis method we have tested using
the simulated spectra generated as described in \S~\ref{sec-spgen};
real spectra can be analyzed in the same manner after continuum
fitting and rescaling.  The basic technique was pioneered by Cowie \&
Songaila (1998; see also Songaila 1998) and improved by Ellison et
al.~(2000) and Schaye et al.~(2000a). In the POD method, the
metal-line optical depth in each pixel is compared to the optical
depth in the pixel corresponding to \HI\ absorption by gas at the same
redshift.  The array of pixel pairs is then binned in \HI\ POD, and
the median metal POD in each bin computed.  For a uniform density
ratio of the metal ion to \HI, and with perfect POD recovery, the
resulting relation would be linear.  In general, the detection of the
metal species will manifest as a correlation between metal POD and
\HI\ PODs, and the change in the slope of this correlation indicates
changes in the ratio of densities (resulting either from changes in
metallicity or ionization balance with density).  The utility of the
POD technique relies upon an understanding of how a given physical
situation maps into the recovered POD relation.  Here we review the
details of the method, and while doing so, compare `recovered' (after
noise, blending, etc.) quantities to `true' (i.e., simulated)
quantities where possible. We give a concise summary of our
recommended implementation of the method in the Appendix.

	Given a QSO redshift $z_{\rm qso}$, we identify the redshift
$z_\beta \equiv (1+z_{\rm qso})(\lambda_{\rm Ly\beta}/\lambda_{\rm
Ly\alpha})-1$ such that absorption by gas in the redshift interval
$[z_\beta,z_{qso}]$ is uncontaminated by Ly$\beta$ absorption.  We
will analyze pixels in the
redshift range $[z_{\rm min},z_{\rm max}]$ with $z_\beta \le z_{\rm
min} < z_{\rm max} \le z_{\rm qso}$. For a transition with rest
wavelength $\lambda_i$ this corresponds to a wavelength range
$\lambda_i(1+z_{\rm min}) \le \lambda \le \lambda_i(1+z_{\rm
max})$.

\subsection{Recovery of $\lya$ optical depths}
\label{sec-optd}

	We derive the $\lya$ (1216\,\AA) optical depth in each pixel
from the flux $F(\lambda)$ using $\tau_{\lya}(\lambda)\equiv-\ln(F)$;
pixels with $\tau_{\lya} < 0$ are discarded.  Pixels are considered
``saturated'' if $F(\lambda) \le N_\sigma\sigma_\lambda$, where
$\sigma_\lambda$ is the rms noise amplitude at the pixel and
$N_\sigma$ is a parameter that we vary in our tests.  For saturated
pixels we can recover a good estimate of the true $\lya$ optical depth
using up to $N_{\rm ho}$ higher-order Lyman lines.  We define
$\tau_{\lya}^{\rm rec} \equiv {\rm min}\left\{\tau_{{\rm Ly}n}f_{{\rm
Ly}\alpha}\lambda_{{\rm Ly}\alpha}/ f_{{\rm Ly}n}\lambda_{{\rm
Ly}n}\right\}$, where $f_{{\rm Ly}n}$ is the oscillator strength of
the $n$th order Lyman line and $\lambda_{{\rm Ly}n}$ is its
rest wavelength (Ly1$= \lya$, Ly2$ = {\rm Ly}\beta$, etc.).  We use
all lines with $1 \le n \le N_{\rm ho}$ that 
lie in the wavelength coverage of the spectrum and for which
$N_\sigma\sigma_n \le F(\lambda_{{\rm Ly}n}) \le 1-N_\sigma\sigma_n$,
where $\sigma_n$ is the noise at $\lambda_{{\rm Ly}n}$. The pixel is
discarded if none of the available higher orders satisfies this
criterion. Taking the minimum optical depth minimizes contamination by
other lines, while the selection criterion picks out strong but
unsaturated lines and minimizes the effects of noise. In particular,
this criterion automatically excludes regions in which the noise is
greater than $\sigma > (2N_\sigma)^{-1}$.

\begin{figure*}
\vbox{ \centerline{ \epsfig{file=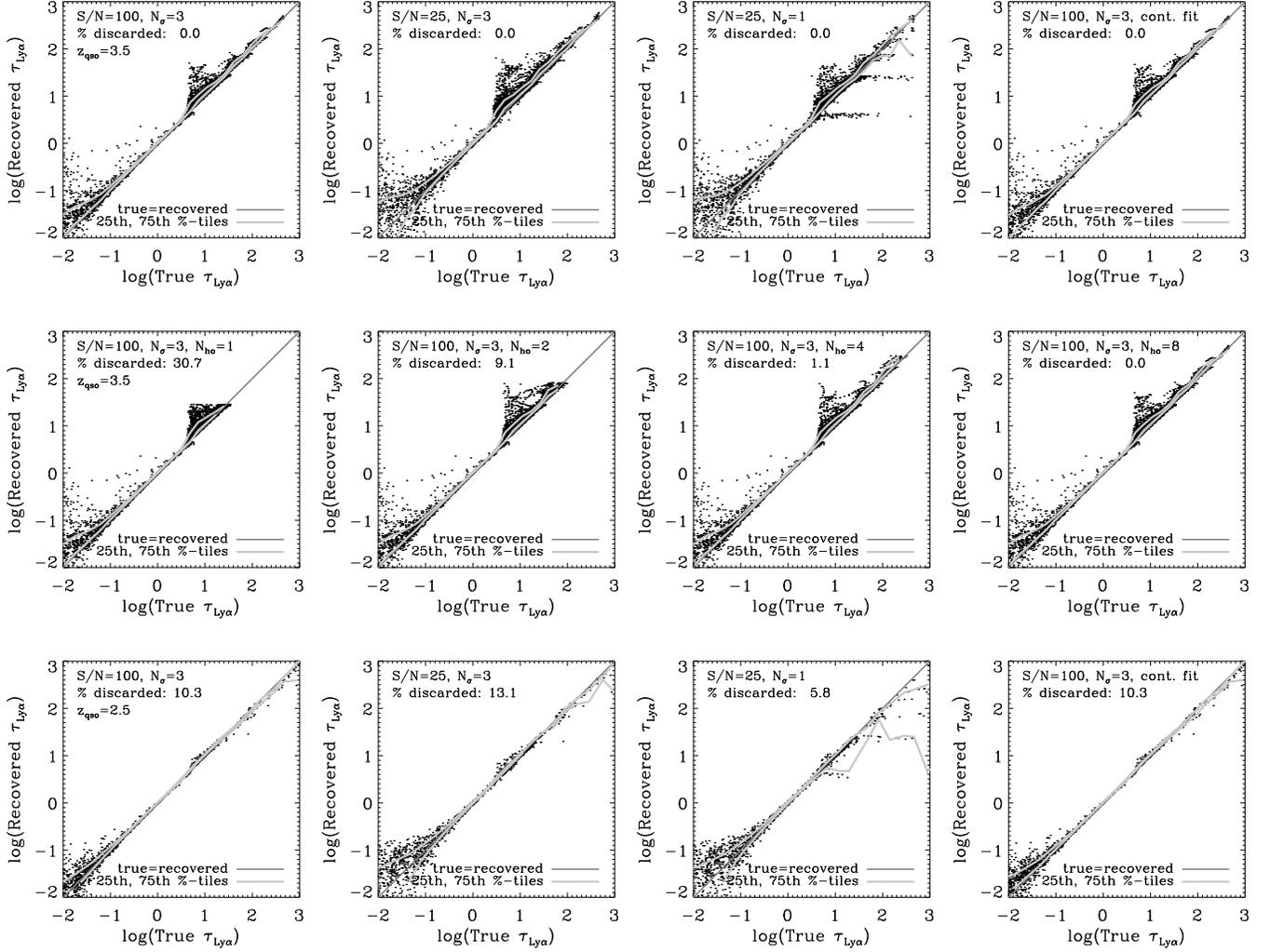,width=18.0truecm}}
\figcaption[]{ \footnotesize Tests of the Ly$\alpha$ optical depth
recovery for saturated pixels, using higher-order transitions.  In
each panel, the `true' (i.e.,\ free of noise, contamination and
instrumental broadening) optical depth $\tau_{\rm true}$ is plotted
against the optical depth $\tau_{\rm rec}$ recovered by the analysis
described in \S~\ref{sec-optd}.  All of the saturated pixels and a
random 10\% of the rest are plotted, and the grey lines indicate 25th
and 75th percentiles of recovered optical depth in each bin of true
optical depth (see text).  For this test we assume $z_{\rm qso}=3.5$
(top and middle panels) or $z_{\rm qso}=2.5$ (bottom panels). The left
two panels of the top and bottom rows show results for 10 higher order
transitions, a threshold $N_\sigma=3$, and $S/N=100$ or $S/N=25$ in
the simulated spectrum.  The third panels from the left on the top and
bottom rows use $N_\sigma=1$.  The rightmost panels on the top and
bottom rows test the sensitivity to errors in the continuum (see text
for details.) The middle row uses $z_{\rm qso}=3.5$, $N_\sigma=3$, and
$S/N=100$, but its four panels show $N_{\rm ho}=1,2,4,$ or 8.  We use
a uniform metallicity $Z=0.01\,Z_\odot$, 0.04\,\AA\ pixels, $6.6\kms$
detector resolution, and a spectrum extending from 3000-7500\,\AA, and
show pixels redward of ${\rm Ly}\beta$ absorption.
\label{fig-rectest}}}
\vspace*{0.5cm}
\end{figure*}

The recovery of saturated $\lya$ optical depth is illustrated in
Figure~\ref{fig-rectest}. For these and other tests we have generated
fiducial simulated spectra as described in \S~\ref{sec-spgen}, with
$z_{\rm qso}=3.5$ (top row) and $z_{\rm qso}=2.5$ (bottom row),
wavelength coverage of 3000-7500~\AA, 0.04~\AA\ pixels and $6.6~\kms$
detector resolution.  A constant metallicity of $Z=0.01\zsol$ is
assumed for the gas. Noise is added to each spectrum to give a
signal-to-noise ratio $S/N$ as described in \S~\ref{sec-spgen}; for
most tests we quote results with both $S/N=25$ and $S/N=100$, to
bracket plausible, high quality observations.

All panels plot the recovered $\lya$ optical depth $\tau_{\rm rec}$
vs.\ the `true' (i.e.\ free of noise, instrumental broadening, and
metal-contamination) value $\tau_{\rm true}$. The grey lines indicate
the 25th(75th) percentiles of $\tau_{\rm rec}$.\footnote{In each bin
the 25th(75th) percentile represents the median of pixels that are
below(above) a curve linearly interpolating the bin medians. They
should not be interpreted as errors on the medians.}  For the left two
panels of the top and bottom rows of Fig.~\ref{fig-rectest} we use
$N_\sigma=3$, $N_{\rm ho}=10$ and $S/N=100$ or $S/N=25$. The optical
depth can be recovered quite reliably for $\tau_{\lya}$ up to several
hundred; the errors are only slightly larger for $S/N=25$ (second
panel of top and bottom rows).  The main error is a slight
over-estimation of optical depth for saturated pixels, due to
contamination (as shown in the bottom four panels, the effect is much
smaller at $z=2.5$ where the $\lya$ forest is less crowded).  Note
that for fixed spectral coverage, the number of higher-order lines
available for reconstructing $\lya$ PODs depends strongly on redshift:
for coverage from $\lambda \ge 3000\,$\AA, all higher order lines are
available only for gas at $z > 2.29$, and only two higher-order lines
are available for gas at $z \approx 2.1$. The middle row shows trials
with the same parameters as the top-left panel, but with $N_{\rm
ho}=1,2,4$, or 8; this illustrates both the effectiveness of the
recovery and the importance of good spectral coverage of the
higher-order lines.

In the third panels from the left (in the top and bottom rows) we show
$N_\sigma=1$ (with $N_{\rm ho}=10$, $S/N=25$); lower $N_\sigma$ gives
less accurate recovery (the `bars' extending to $\tau_{\rm true} >
\tau_{\rm rec}$ are caused by noise), though less pixels are discarded
as having no reliable optical depth estimate.  Even for $N_\sigma=3$
the number of discarded pixels is quite small (none at $z=3.5$ and
$\sim 10\%$ at $z=2.5$), so we set $N_\sigma=3$ in all subsequent
trials unless otherwise noted.\footnote{For a realistic noise array in
which noise increases in the far blue, somewhat more pixels are
discarded because coverage of the higher-order lines is
lost. Particularly at $z < 3$, this might call for a lower value of
$N_\sigma$ to maintain good statistics at large $\tau$ values.}

The rightmost panels (of the top and bottom rows) show the effects of
errors in the continuum fit, estimated using the following procedure.
We divide the spectrum into 90\,\AA\ bins with centers $\lambda_k$,
and find for each bin the median flux $\bar f_k$.  We then interpolate
a spline across the $\bar f_k$, and flag all pixels that are at least
$N_\sigma^{\rm cf}\sigma$ below the interpolation. The medians are
taken again using the unflagged pixels and the procedure is repeated
until the fit converges.  The fluxes and errors are then scaled by the
fitted continuum.  The accuracy of the fit depends upon $N_\sigma^{\rm
cf}\sigma$; fitting in highly-absorbed regions is more accurate with
low $N_\sigma^{\rm cf}$ (because more low-flux pixels are rejected),
whereas relatively unabsorbed regions are better fit with high
$N_\sigma^{\rm cf}$.  We find that using $N_\sigma^{\rm cf}=1$ when
analyzing \CIV\ and using $N_\sigma^{\rm cf}=0.5$ for \OVI\ gives an
acceptable continuum fitting error in the metal line regions (and
either value gives unimportant errors in the $\lya$ region).
Comparison between the original and scaled spectra then shows that for
$z_{\rm qso}=3.5$ the induced errors are $\sim +0.23$ -- $+1.3\%$ in
the $\lya$ region, $\sim +1.4\%$ in the \OVI\ region and $\sim
+0.06\%$ in the \CIV\ region; for $z_{\rm qso}=2.5$ the errors are
$\sim -0.4$ -- $+0.4\%$, $\sim 0.1\%$, and $\sim -0.1\%$ respectively.
As shown in the figure, this continuum error does not have any
significant effect on the recovered $\lya$ optical depths; it will
however, be more important in recovering the metal-line optical depths
as discussed below.

\subsection{Recovery of metal line optical depths}

	Having obtained for each pixel a good estimate of the $\lya$
optical depth (which, as shown below in \S~\ref{sec-hiinterp} can be
converted into a density of the absorbing gas), we now turn to the
metal-line optical depths for various species, which we will use to
recover information about the metallicity of the absorbing gas.

	For each pixel with $z_{\rm min} \le z \le z_{\rm max}$ we
find the wavelength corresponding to absorption by some metal species,
$\lambda = \lambda_i (1+z)$, where $\lambda_i$ is the rest wavelength
of the metal line transition; in this study we investigate the
recovery of the species \OVI, \CIV, \NV\ and \SiIV, concentrating on
the first two.  The species we consider all appear in doublets
(labeled here with subscripts `I' and `II'), with the stronger
transition (`I') twice as strong as the weaker (i.e.,
$f_{i,I}\lambda_{i,I}=2f_{i,II}\lambda_{i,II}$ where $f$ is the
oscillator strength). We then compute a metal line optical depth
associated with each pixel of $\tau_i(z)={\rm
min}(\tau_{i,I}(z),2\tau_{i,II}(z))$, i.e., the minimum of the optical
depth of the stronger and twice the optical depth of the weaker
component of the doublet.\footnote{Self-contamination can be corrected
(see below) in regions (such as \CIV) where it is more important than
contamination by other lines, and in this case the minimum should {\em
not} be taken.}  Taking the minimum decreases contamination effects,
which are particularly important for \NV\ and \OVI. Different from
previous work, we do not discard pixels for which neither component is
well-detected. Instead, we assign an optical depth
$\tau_i(z)=-\ln(N_\sigma\sigma_{\lambda_{i,I}})$ (if the line is
saturated) or $\tau_i(z)=\tau_{\rm min}\equiv 10^{-5}$ (if the optical
depth is negative). Because there is little scatter in metal POD at
high $\lya$ POD, the former prescription causes $\tau_Z(\tau_{\rm HI})$
to flatten smoothly to the highest detectable value.  The latter
prescription allows pixels with very low optical depth to correctly
influence the median POD (the number of well-detected versus
poorly-detected pixels contains information about the metallicity) as
long as $\tau_{\rm min}$ is smaller than the median in any bin, which
holds for $\tau_{\rm min} \la 10^{-5}$.

\begin{figure*}
\vbox{ \centerline{ \epsfig{file=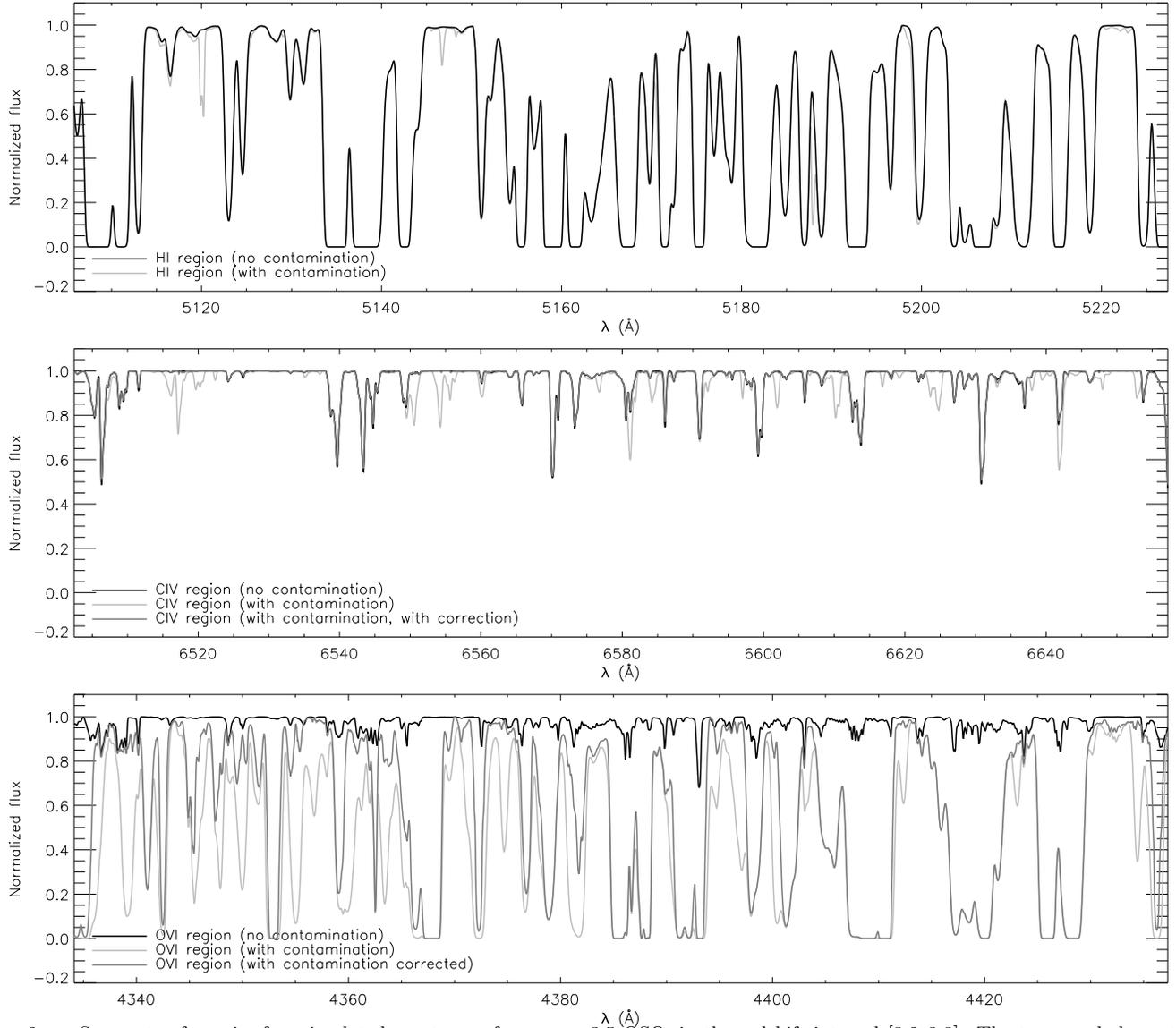,width=18.0truecm}}
\figcaption[]{ \footnotesize Segments of a noise-free simulated
spectrum of a $z_{\rm qso}=3.5$ QSO, in the redshift interval
$[3.2,3.3]$. The top panel shows the Ly$\alpha$ region
with metal contamination included (light line) and not included (dark
line).  The middle panel shows the corresponding region for
\CIV. The black line includes only the stronger doublet component, and
is not contaminated by other metals (the black line is nearly
invisible because the dark grey line falls almost exactly on top of
it). The light line includes both doublet 
components and contamination (by silicon). The dark grey line line
shows the (contaminated) spectrum corrected for \CIV\ {\em
self}-contamination (see text). The lower panel's dark line is the
\OVI\ region, with all contaminants turned off.  The light line shows
the spectrum with \HI\ and other metal absorption included, and the
darker line shows the spectrum after removal (see text) of the
higher-order Lyman lines. A uniform metallicity of $Z=0.01\zsol$ was
used for all panels.
\label{fig-sampspec}}}
\vspace*{0.5cm}
\end{figure*}

\begin{figure*}
\vbox{ \centerline{ \epsfig{file=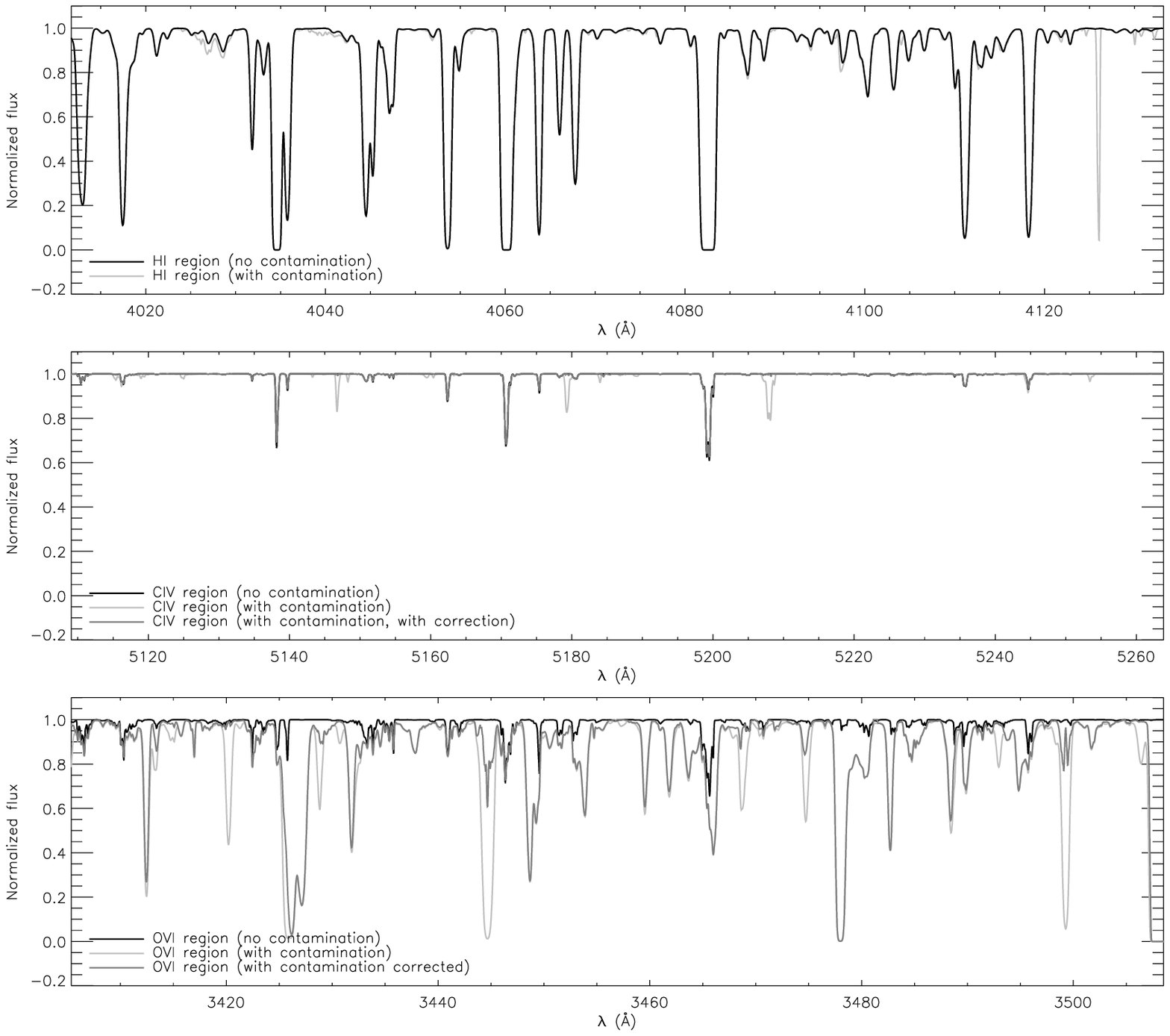,width=18.0truecm}}
\figcaption[]{ \footnotesize Segments of a noise-free simulated
spectrum of a $z_{\rm qso}=2.5$ QSO, in the redshift interval
$[2.3,2.4]$ in Ly$\alpha$.  Plotted lines are as in
Fig.~\ref{fig-sampspec}.
\label{fig-sampspec_z2.5}}}
\vspace*{0.5cm}
\end{figure*}

	Given $z_{\rm min} > z_\beta$, the four chemical species we
consider suffer very different levels of contamination from hydrogen
lines. Carbon IV is contaminated only by other metals and by its own
doublet, and \SiIV\ is similarly uncontaminated through most of its
available wavelength range. On the other hand, \OVI\ is contaminated
by several Lyman lines including $\lya$, and the wavelength range for
\NV\ overlaps almost completely with that of Ly$\alpha$ (although
there is a tiny segment with $\delta z\sim 0.1$ for which \NV\ lies
red-ward of the QSO's $\lya$ emission line and is thus relatively
uncontaminated).  The importance of contamination can be seen in
Figs.~\ref{fig-sampspec} and~\ref{fig-sampspec_z2.5}, which show, for
$z_{\rm qso}=3.5$ and $z_{\rm qso}=2.5$ respectively, examples of a
small spectrum segment in $\lya$ region (top panel) and the
corresponding \CIV\ (middle panel) and \OVI\ regions (bottom panel).
All three panels are plotted both with and without contamination by
Lyman series and metal lines.

A significant fraction of the contamination in the OVI region is due
to higher-order Lyman transitions, and this contamination can be partially
removed using the (accurate) estimate of the $\lya$ optical depth that
has been computed {\em using} those higher-order lines as
described in \S~\ref{sec-optd}.  To do this, for each pixel with
wavelength $\lambda$ in the \OVI\ region, we cycle through $N_{\rm
corr}$ higher-order lines.  For each line $i=1,\ldots, N_{\rm corr}$
we check the recovered $\lya$ optical depth $\tau_{\rm Ly\alpha}$ at
wavelength $\lambda' = (\lambda_{\rm Ly\alpha}/\lambda_{{\rm
Ly}i})\lambda$, and subtract the optical depth $\tau_{{\rm
Ly}\alpha}(\lambda')\left(f_{{\rm Ly}i}\lambda_{{\rm Ly}i}/f_{{\rm
Ly}\alpha}\lambda_{{\rm Ly}\alpha}\right)$.  As can be seen in
Figs.~\ref{fig-sampspec} and~\ref{fig-sampspec_z2.5}, this correction
is quite helpful.  For a given QSO redshift, the importance of this
correction increases (as does \HI\ contamination itself) with
decreasing redshift of the absorbing gas, because progressively more
higher-order Lyman lines contribute to the contamination.

\begin{figure*}
\vbox{ \centerline{ \epsfig{file=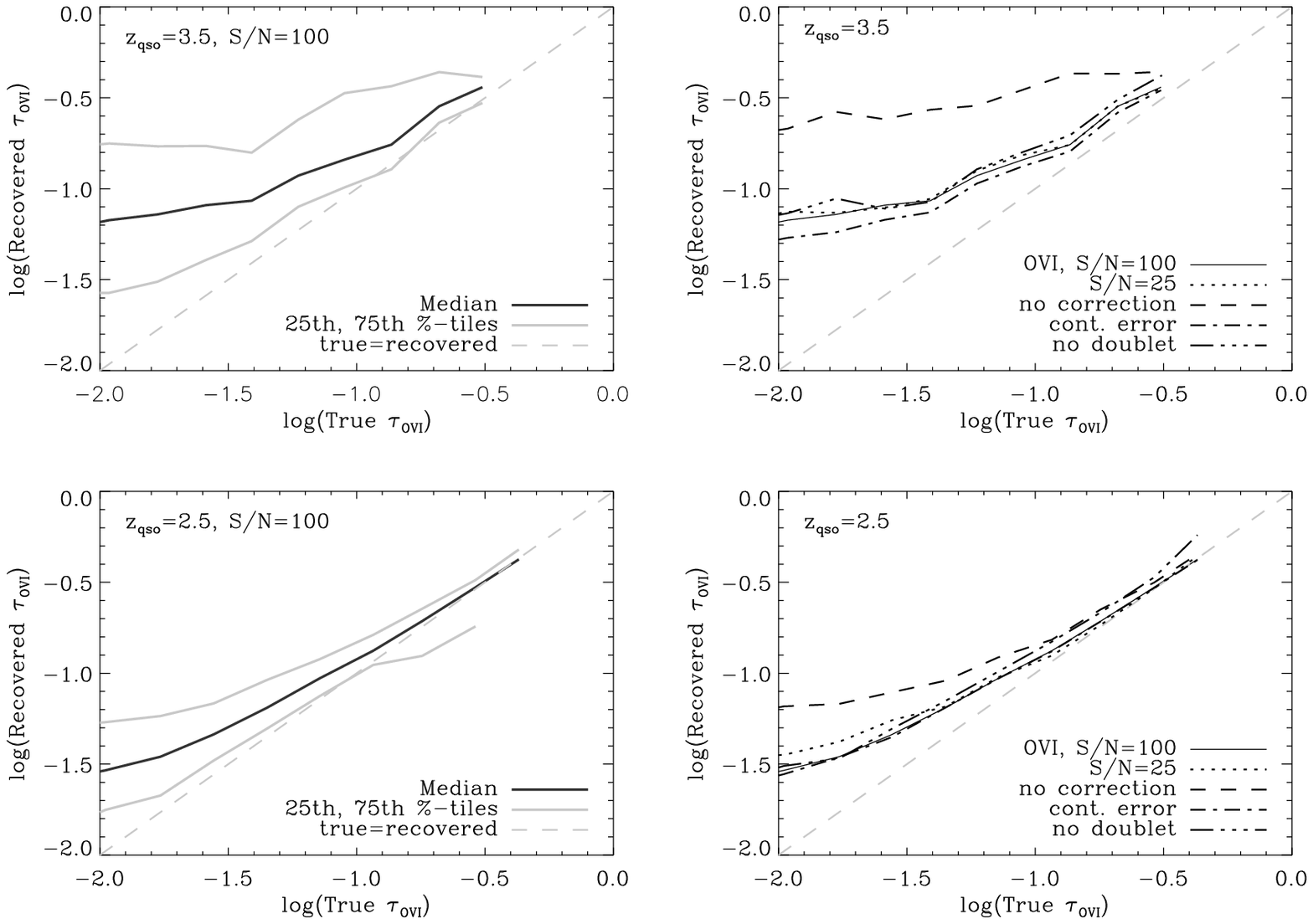,width=18.0truecm}}
\figcaption[]{ \footnotesize Tests of the recovery of the `true' OVI
pixel optical depth from the simulated spectra.  The left panels show
the median, 25th and 75th percentiles in recovered optical depth,
versus the noise-free, perfect resolution and uncontaminated optical
depth.  The top panels assume $z_{\rm qso}=3.5$ and $[z_{\rm
min},z_{\rm max}]=[3.1,3.5]$; the bottom assume $z_{\rm qso}=2.5$ and
$[z_{\rm min},z_{\rm max}]=[2.1,2.5]$.  The left panels use $S/N=100$.
The right panels (plotting only the medians) show three
variations: $S/N=25$ (dotted lines), without removal (see text) of
higher-order Lyman lines from the OVI region (dashed lines), and with
continuum fitting errors included (dot-dashed lines).
All panels use $N_\sigma=3$ and a uniform metallicity of 1\% solar.  
\label{fig-recztest_ovi}}}
\vspace*{0.5cm}
\end{figure*}

Figure~\ref{fig-recztest_ovi} shows how well the true \OVI\ optical
depths can be recovered, using a simulated spectrum.  Each panel plots
the median binned recovered \OVI\ POD versus the true one (i.e.\ free
of noise, contamination, and instrumental broadening).  Top panels
are for $z_{\rm qso}=3.5$ while bottom panels show $z_{\rm qso}=2.5$.
The left panels show the median and 25th and 75th percentiles in
recovered $\tau_{\rm OVI}$ for our fiducial model (note that these
percentiles are {\em not} errors in the median). There is a clear
correlation between the recovered and true \OVI\ optical depths and
the recovery works particularly well at $z_{\rm qso}=2.5$ where there
is far less contamination. We have assumed a uniform metallicity; the
correlation signal would be steeper if the metallicity were to
increase with density.  The panels on the right show the median curves
with more noise ($S/N=25$), without correcting for contamination from
higher-order Lyman lines, or with errors in the continuum fit
included. Note that in both cases the continuum fitting errors work in
the opposite sense from most others, {\em decreasing} the \OVI\
optical depths; this effect is discussed further in
\S~\ref{sec-restest}.  Overall, recovery of \OVI\ PODs is imperfect, even
for fairly large $\tau_{\rm OVI}$, but in \S~\ref{sec-invert} we shall
discuss how the simulations can be used to correct for this imperfect
recovery.

Self-contamination can be corrected for if it dominates other
contamination, and we implement this correction for \CIV. For each
pixel at wavelength $\lambda$ in the \CIV\ region we check the optical
depth (computed \emph{without} taking the minimum of the doublet) at
the wavelength $\lambda\lambda_I/\lambda_{II}$ where $\lambda_I$ and
$\lambda_{II} > \lambda_I$ are the rest wavelengths of the \CIV\
doublet. Half this optical depth (the theoretical strength of the
second doublet component) is subtracted from the initial POD estimate,
and the process is iterated (and converges after about 5 iterations).

The \CIV\ self-contamination correction works very well, but can
enhance the effect of other strong contaminating lines that may be
present in the \CIV\ region of observed spectra (e.g., \ion{Mg}{2}),
by erroneously subtracting a doublet component as if they were \CIV\
absorption.  If such lines are not removed by hand, it is therefore
important to remove these contaminants; this can be done automatically
by discarding all pixels in the \CIV\ region with optical depth
$\tau(\lambda)$ satisfying
$$
\exp(-\tau)+3\sigma < \exp[-2\tau(\lambda\lambda_{II}/\lambda_{I})-\tau(\lambda\lambda_{I}/\lambda_{II})/2],
$$
where $\sigma$ is the r.m.s. noise of the pixels at $\lambda$,
$\lambda\lambda_I/\lambda_{II}$, and
$\lambda\lambda_{II}/\lambda_{I}$.  Such pixels have too much
absorption to be the sum of any doublet (with a primary component of twice the
stength at $\lambda\lambda_I/\lambda_{II}$ and a primary component (with a
doublet of half the strenth at $\lambda\lambda_{II}/\lambda_{I}$), and
are very likely contaminated. The self-contamination correction should
then be done.  We have tested this procedure by setting the flux $F=0$
in a random fraction $f$ of the pixels in the \CIV\ region.  For $f=0$
the contaminant removal does not change the recovered $\tau_{\rm
CIV}$, and with both contaminating lines and self-contamination
removed as described, the recovery is not significantly affected for
$f \la 0.2$ (if {\em no} contaminant removal is done, but the minimum
of the doublet components is taken, then the recovery is still only
significantly affected for $f \ga 0.1$) This robustness of median
statistics is, of course, one of the advantages of the median POD
method.

\begin{figure*}
\vbox{ \centerline{ \epsfig{file=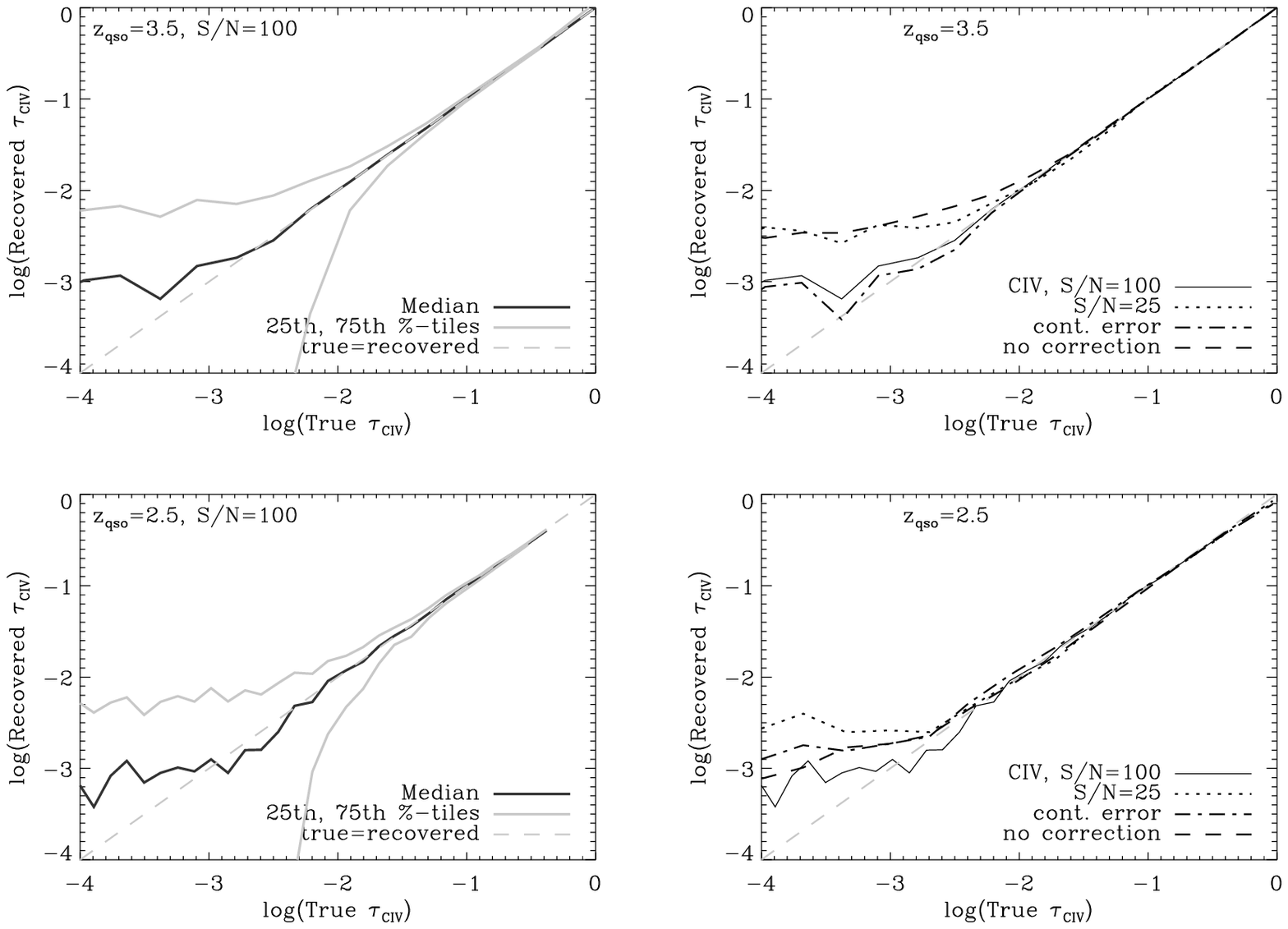,width=18.0truecm}}
\figcaption[]{ \footnotesize As in previous figure, but for CIV.
Here, the top panels assume $z_{\rm qso}=3.5$ and $[z_{\rm min},z_{\rm
max}]=[2.8,3.5]$; the bottom panels assume $z_{\rm qso}=2.5$ and $[z_{\rm
min},z_{\rm max}]=[1.95,2.5]$.  The three variations shown are:
$S/N=25$ (dotted lines), with continuum fitting errors
included (dot-dashed lines), and without correction for
self-contamination (but with the doublet minimum taken) (dashed lines).
\label{fig-recztest_civ}}}
\vspace*{0.5cm}
\end{figure*}

Figure~\ref{fig-recztest_civ} shows tests of the CIV recovery.  As
shown in the left panel, the recovery is nearly perfect down to
$\tau_{\rm CIV} \sim 10^{-3}$ at both redshifts.  In the right panel,
the dotted and dot-dashed curves show how the recovery becomes
slightly less accurate with $S/N=25$ or with a continuum fitting error
included.  Here, continuum fitting errors arise more due to noise than
to significant absorption and do {\em not} tend to systematically
reduce optical depth\footnote{This is only true if $N_\sigma^{\rm cf}$
is carefully chosen; otherwise the continuum fitting can
systematically increase or decrease the optical depth in a relatively
empty region.}; instead they act as an additional source of random
error that reduces the accuracy of the $\tau_{\rm CIV}$ recovery.  The
dashed line shows the curve without the self-contamination correction
(but with the minimum of the doublet components taken); this shows
that for $z_{\rm qso}=3.5$ and our assumed metallicity, self
contamination is the limiting factor in the \CIV\ POD recovery if not
corrected for (as was the case in previous studies).

\section{Testing and interpreting the POD method}
\label{sec-interp}

The previous section described the recovery of metal and $\lya$ PODs
from a simulated (or observed) spectrum, and tested the accuracy of
this recovery process.  In this section we use our simulations to draw
connections between the recovered PODs and physical quantities such as
gas density and metallicity.

\subsection{Interpreting \HI\ PODs}
\label{sec-hiinterp}

\begin{figure*}
\vbox{ \centerline{ \epsfig{file=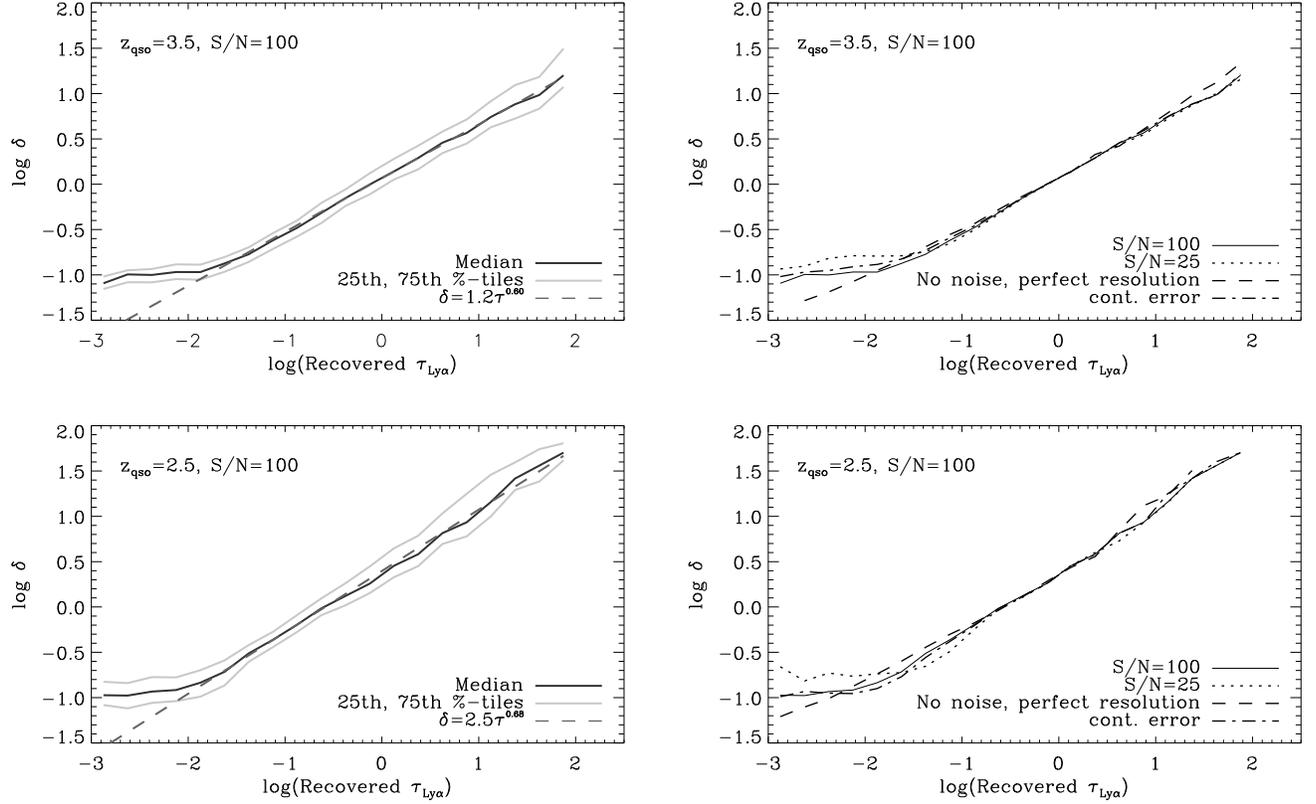,width=18.0truecm}}
\figcaption[]{ \footnotesize Ly$\alpha$ optical depth weighted overdensity
of gas responsible for the absorption in each pixel, versus the
recovered Ly$\alpha$ POD in that pixel, binned in $\tau_\lya^{\rm
rec}$. Left panels plot the median and 25th and 75th percentiles of gas
overdensities corresponding to a given pixel optical depth; top panels
show $[z_{\rm min},z_{\rm max}, z_{\rm qso}]=[2.8,3.5,3.5]$ while
bottom panels show $[z_{\rm min},z_{\rm max}, z_{\rm
qso}]=[1.95,2.5,2.5]$. Left panels show the fiducial
$S/N=100$ model.  A least-absolute-deviation power law fit is also
included and gives the relations $\delta=1.2\tau_{\lya}^{0.60}$
($z=3.5$) and $\delta=2.5\tau_{\lya}^{0.68}$ ($z=2.5$).  Right panels
(showing only medians) show trials with $S/N=25$, with continuum
errors included, and with no noise and infinite detector resolution.
The optical depths are recovered using up to 10
higher order lines, and $N_\sigma=3$. 
\label{fig-tauvdel}}}
\vspace*{0.5cm}
\end{figure*}

	Previous studies -- both numerical (e.g., Zhang et al. 1998,
Dav\'e et al. 1999; Croft et al. 1997,1998) and analytical (Schaye
2001) -- have shown that there is a fairly tight relation between the
column density or optical depth of $\lya$ absorption, and the density
of the gas responsible for the absorption. Figure~\ref{fig-tauvdel}
confirms that a tight relation between optical depth and gas density
holds on a pixel-by-pixel basis. All panels plot the median $\lya$
optical depth weighted overdensity versus the recovered $\tau_\lya$
for $z_{\rm qso}=3.5$ (top panels) and $z_{\rm qso}=2.5$ (bottom
panels), using $N_{\rm ho}=10$ and $N_\sigma=3$.  The left panels show
the median overdensity (dark lines) and the 25th and 75th percentiles
(light lines) for $S/N=100$. The relation is very close to a
power-law, except at very low optical depth, $\tau_{\lya} \la
10^{-2}$, where it flattens off. Least absolute deviation fits (dashed
lines) yield power-law indices of 0.60 ($z=3.5$) and 0.68
($z=2.5$). Neglecting redshift space distortions, theory predicts a
power-law relation of $\tau_{\lya} \propto \delta^{2.76-0.76\gamma}$,
where the gas temperature obeys $T=T_0\delta^{\gamma-1}$ (Hui \&
Gnedin 1997) with $ 1 \la \gamma \la 1.6$. This predicts then
$\delta\propto \tau_{\lya}^\alpha$ with $0.5 \la \alpha \la 0.65$, in
agreement with our results. The existence and tightness of this
relation shows that for reasonable observational errors the density of
absorbing gas can be recovered to within $\sim \pm 50\%$ about half of
the time for {\em each} pixel, and very reliably on a statistical
basis.  This will allow us to readily translate information gleaned
using pixel optical depths into information regarding gas of specific
density (though with some uncertainty coming from uncertainty
regarding the ionizing background).  The right panels show medians
only, for $S/N=25$ (dotted lines), with perfect detector resolution
and no noise (dashed lines), and with continuum fitting errors
included (dot-dashed).  These curves are all quite close except for
optical depths lower than the errors (mainly due to noise).

\subsection{Interpreting metal PODs}
\label{sec-zinterp}

\begin{figure*}
\vbox{ \centerline{ \epsfig{file=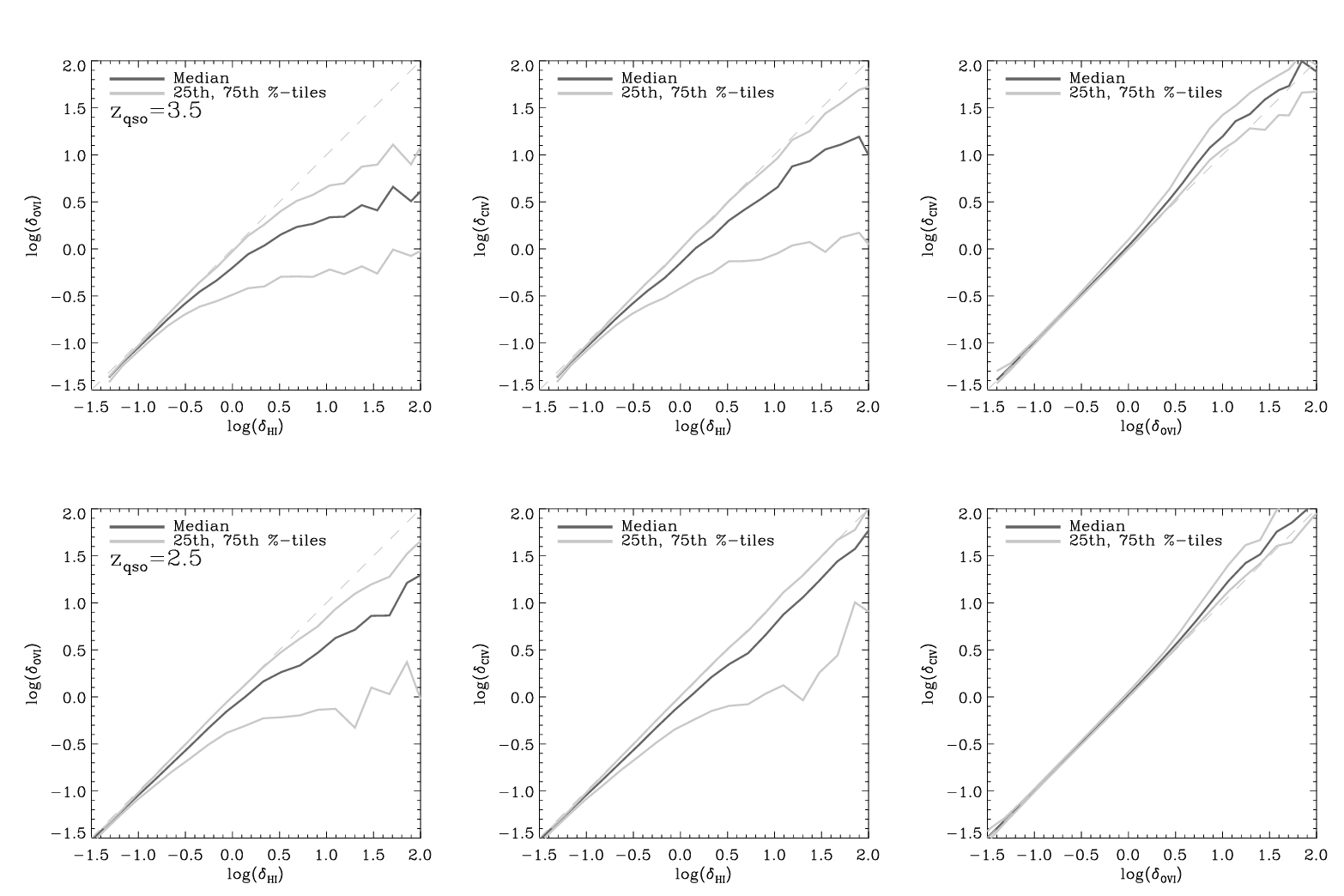,width=18.0truecm}}
\figcaption[]{\footnotesize Comparison of overdensity weighted by
$\tau_{\lya}$ versus that weighted by $\tau_{\rm OVI}$ (left panels)
or $\tau_{\rm CIV}$ (middle panels), assuming $z_{\rm qso}=3.5$ (top
panels) or $z_{\rm qso}=2.5$ (bottom panels).  The right panels plot
the two metal optical depth weighted densities against each other.
Metallicity is uniform at 1\% solar, and there is no noise in the
spectrum.
\label{fig-wgttest}}}
\vspace*{0.5cm}
\end{figure*}

Given the tight relation between POD and gas density, the ratio of
metal to $\lya$ PODs is related to the ratio of metal ion to \HI\
density (subject to the errors in the recovery discussed in
\S~\ref{sec-optd}), which is in turn related (by an ionization
correction) to the gas metallicity. One potential pitfall in this
chain of inference is that both ionization effects and the difference
in the thermal widths of metal and hydrogen absorption may lead to
slight offsets between the $\lya$ and metal lines (see Ellison et
al. 2000 for some discussion). It is therefore important to test
whether the absorption by metals and hydrogen arises in the {\em same
gas} on a pixel-by-pixel basis.  Figure~\ref{fig-wgttest} tests
whether the \OVI\ and \CIV\ absorption in a given pixel is due to gas
of the same density as the gas giving rise to the $\lya$
absorption. The left and middle panels plot the gas overdensity
weighted by the \CIV\ and \OVI\ optical depth respectively (denoted by
$\delta_{\rm OVI}$ and $\delta_{\rm CIV}$) against the $\lya$-optical
depth weighted density $\delta_{\rm HI}$ for $z_{\rm qso}=3.5$ (top
panels) and $z_{\rm qso}=2.5$ (bottom panels). The correlation is good
but not perfect. For high $\delta_{\rm HI}$ the metal (and in
particular the \OVI) optical depth weighted densities are low compared
with $\delta_{\rm HI}$. This effect is mainly due to the greater
thermal width of the hydrogen lines (tests show that the correlation
becomes very tight if the detector resolution approaches the typical
$\lya$ line width or if the metals are given atomic weight unity so
that their lines are as broad as $\lya$).  This discrepancy will
always be unimportant for low ($\log(\delta) \la 0.5$) densities, but
could potentially smear the $Z(\delta)$ relation for high
densities. However, as we show in \S~\ref{sec-invert}, this does not
appear to be a significant effect, at least not for the density range
over which we can measure the metallicity ($\delta_Z \la 10$).

If the gas temperature is close to $\sim 3\times 10^5$\,K, then
collisional ionization becomes important for \OVI, and \OVI\ is
therefore often regarded as a probe of hot gas. To reach such high
temperatures, the gas needs to be shock-heated. At low redshift ($z\la
1$), gravitational accretion shocks may well heat a
significant fraction of the baryons to temperatures $T\ga 10^5~{\rm
K}$, but the fraction of gravitationally heated gas is predicted to be
much smaller at the redshifts of interest here ($z\ga 2$) (e.g., Cen
\& Ostriker 1999). Indeed, we find that the fraction of shock-heated
gas in our simulation is small enough for collisional ionization to be
negligable. However, it should be noted that our simulations contain
only gravitational shocks. If shocks from galactic winds keep a
significant fraction of the high-$z$ IGM in a hot phase, then our
simulations would underestimate the fraction of collisionally ionized
\OVI. If this were the case then the inferred median \OVI\ optical depth
would of course still be correct. However, as mentioned in
\S~\ref{sec-intro}, the (incorrect) assumption of pure
photoionization would then lead to inacurate conclusions concerning
the oxygen abundance. 

\subsection{Testing the method}
\label{sec-restest}

\begin{figure*}
\vbox{ \centerline{ \epsfig{file=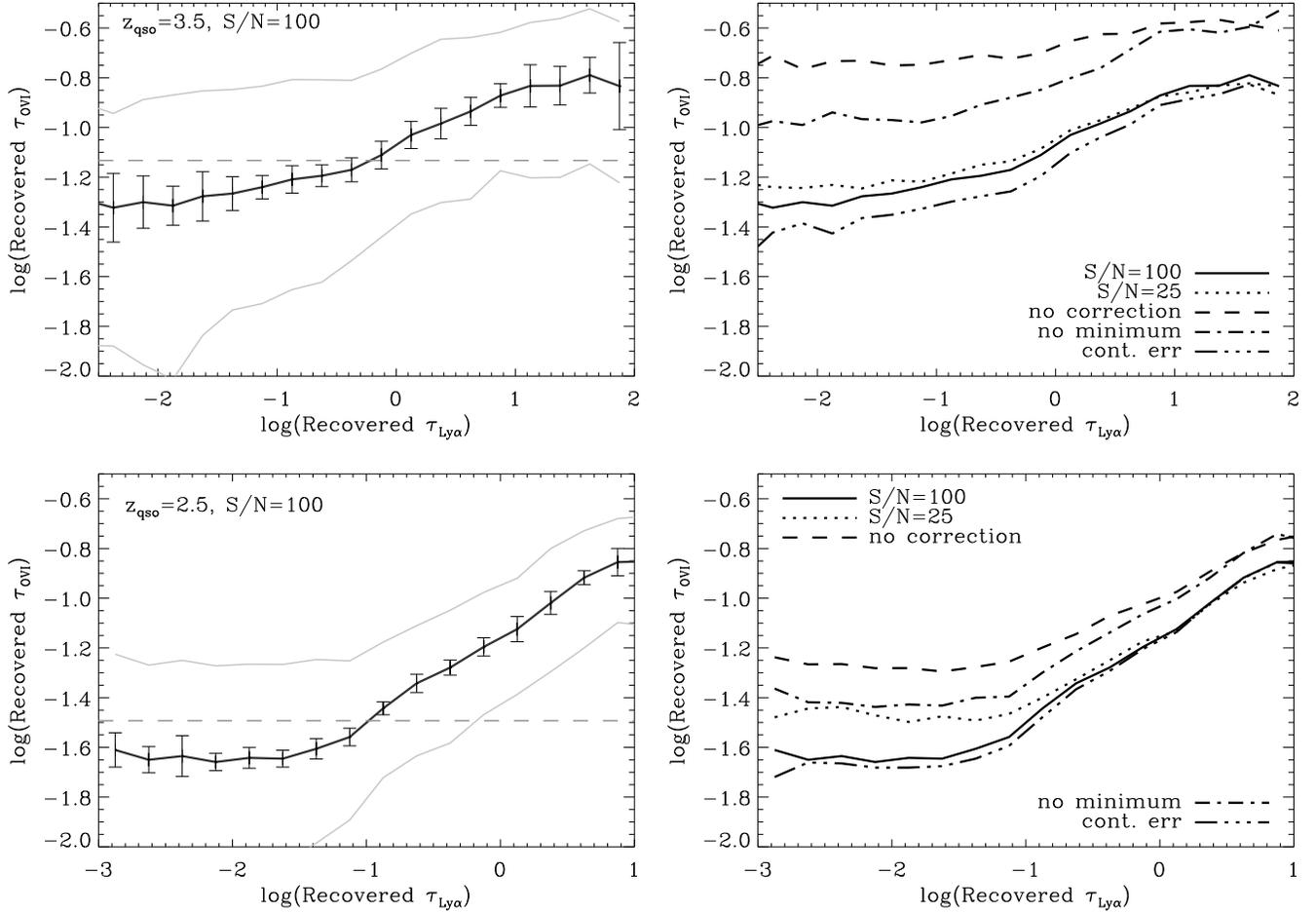,width=18.0truecm}}
\vskip0.15in
\figcaption[]{ \footnotesize Recovered OVI optical depth vs.\
recovered $\lya$ optical depth for $z_{\rm qso}=3.5$ (top panels) and
$z_{\rm qso}=2.5$ (bottom panels).  Left panels assume a uniform
metallicity $Z=0.01\zsol$, $z_{\rm min}=3.1$, $z_{\rm max} = z_{\rm
qso}$, $N_\sigma=3$, and $S/N=100$.  The median and 25th and 75th
percentiles are averaged over ten independent spectra, and the error
bars represent the standard deviations of the medians about the mean
median.  The horizontal dashed lines indicate the median $\tau_{\rm
OVI}$, irrespective of $\tau_{\lya}$.  Right panels show mean
medians for the fiducial model (solid lines) and various other models:
with $S/N=25$ (dotted lines), without correction for higher-order
Lyman lines (dashed lines), without taking the minimum of the doublet,
(dot-dashed line), or with continuum fitting error included
(triple-dot-dashed lines).
\label{fig-odtest_ovi}}}
\vspace*{0.5cm}
\end{figure*}

\begin{figure*}
\vbox{ \centerline{ \epsfig{file=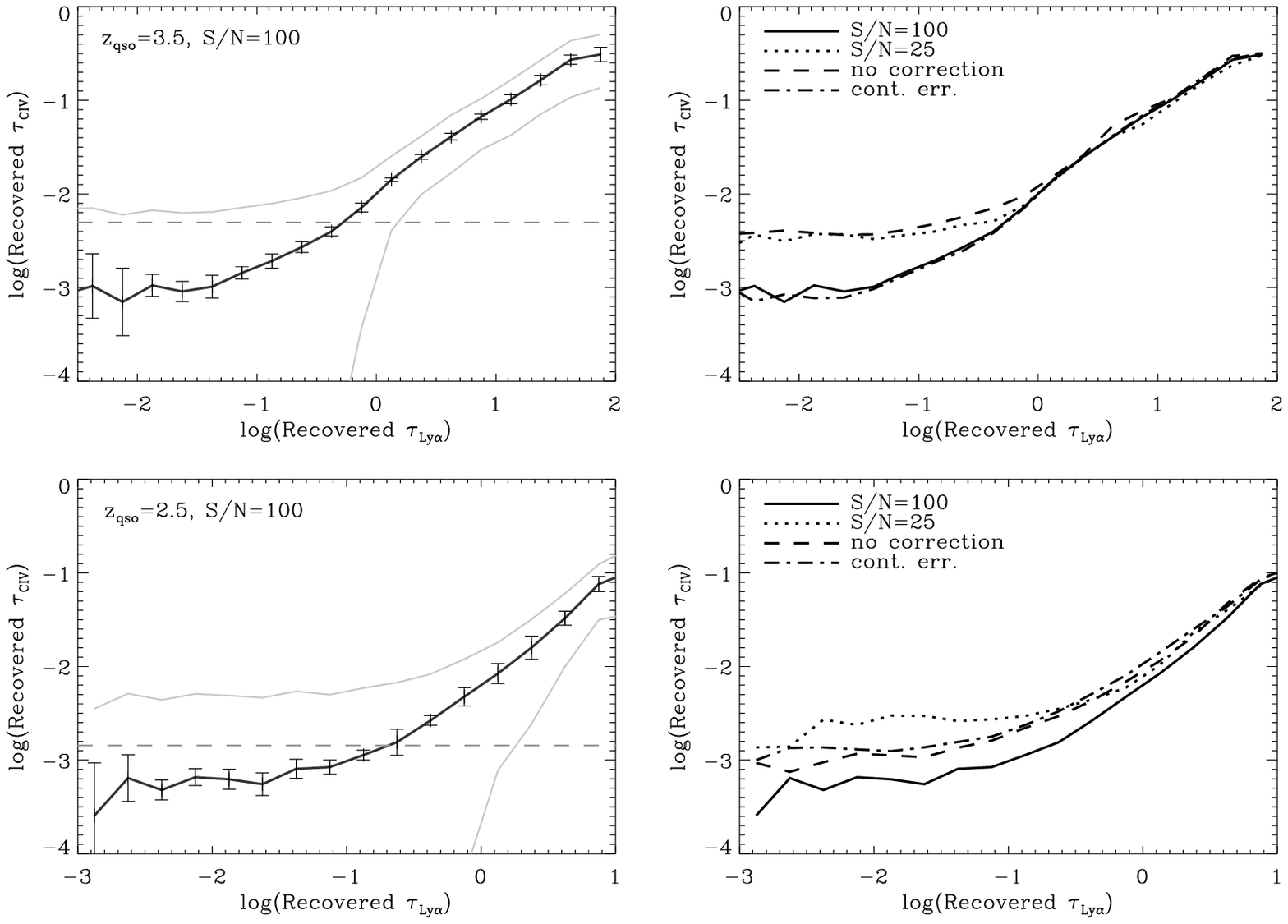,width=18.0truecm}}
\vskip0.15in
\figcaption[]{Recovered CIV optical depth vs.\ recovered $\lya$
optical depth for $z_{\rm qso}=3.5$ (top panels) and $z_{\rm qso}=2.5$
(bottom panels).  Left panels assume uniform metallicity
$Z=0.01\zsol$, $z_{\rm min}=3.1$, $z_{\rm max} = z_{\rm qso}$,
$N_\sigma=3$, and $S/N=100$.  The median and 25th and 75th percentiles
are averaged over ten independent spectra, and the error bars
represent the standard deviations of the medians about the mean
median.  The horizontal dashed lines indicate the median $\tau_{\rm
CIV}$, irrespective of $\tau_{\lya}$.  Right panels show mean
medians for the fiducial model (solid lines) and various other models:
with $S/N=25$ (dotted lines), without self-contamination correction
(dashed lines), or with continuum fitting errors included (dot-dashed
lines).  \footnotesize
\label{fig-odtest_civ}}}
\vspace*{0.5cm}
\end{figure*}

	Having shown that the recovered $\lya$ optical depth 
is related to the gas density and that the metal absorption arises in
gas with roughly the same density as the gas responsible for the
hydrogen absorption, we will now investigate the relation between the
measured $\tau_Z^{\rm rec}(\tau_\lya^{\rm rec})$ and the metallicity
as a function of the gas density. In this section we will  
investigate the effects of variations in the recovery procedure and
the metallicity $Z(\delta)$ on the correlation between metal and
$\lya$ optical depth. Figures~\ref{fig-odtest_ovi}
and~\ref{fig-odtest_civ} show the 
basic results of the POD method applied to simulated spectra with
uniform metallicity $Z=0.01~Z_\odot$.  Figure~\ref{fig-odtest_ovi}
applies the method to \OVI, using fiducial parameters ($N_{\rm
ho}=10$, $N_{\rm corr}=4$, and $N_\sigma=3$).  The top panels use
$[z_{\rm min},z_{\rm max}=z_{\rm qso}]=[3.1,3.5]$ while the bottom panels
use $[z_{\rm min},z_{\rm max}=z_{\rm qso}]=[2.1,2.5]$.  Each panel
plots $\tau_{\rm OVI}^{\rm rec}$ vs.\ $\tau_{\lya}^{\rm rec}$, binned
into 20 bins of 
$\log(\tau_{\lya}^{\rm rec})$ in the range $[-3,3]$.  Dark lines indicate
medians in each bin, while light lines (in left panels) signify 25th
and 75th percentiles. The medians and percentiles are calculated
separately for each of ten independent spectra, and the average of the
medians and percentiles are plotted; the error bars about the medians
in the left panels represent the standard deviation of the medians
about the mean median in each bin.

	The correlation between $\tau_{\rm OVI}^{\rm rec}$ and
$\tau_{\lya^{\rm rec}}$, which is clearly evident in the $z_{\rm
qso}=2.5$ case for $\tau_{\lya}^{\rm rec} \ga 0.03$, indicates the
detection of \OVI\ in gas of overdensity $\delta \ga 0.2$ (using
Fig.~\ref{fig-tauvdel}); at $z_{\rm qso}=3.5$ the correlation is
significant at $\tau_{\lya}^{\rm rec} \ga 0.1$, or $\delta \ga 0.5$.
The dashed horizontal line in each panel on the left represents the
median $\tau_{\rm OVI}^{\rm rec}$, irrespective of $\tau_{\lya}^{\rm
rec}$. If no \OVI\ were present, all data points would (and tests show
do) lie near this horizontal line.  Note that unless \OVI\ contributes
only negligibly to the mean optical depth in the metal region, this
line is {\em not} a ``reference level'' (or detection limit) that
gives the contribution of noise and other errors in the absence of
detected metal absorption; the latter would correspond to the constant
$\tau_{\rm OVI}^{\rm ref}$ to which the curve asymptotes as
$\tau_{\lya}^{\rm rec} \rightarrow 0$.  One might hope to subtract
such a ``reference level'' (as in Cowie \& Songaila 1998) to see more
clearly where metals are detected, but this requires a very accurate
estimate of $\tau_{\rm OVI}^{\rm ref}$, and will generally lead to
negative optical depths and large errors at low $\tau_{\lya}$.  Thus,
for practical purposes we prefer {\em not} to subtract the reference
level, and instead recover metallicity information as described below
in \S~\ref{sec-invert}.

The right panels show the effects of different assumptions or
parameters in the analysis.  The dashed lines, where no correction for
higher-order Lyman lines in the \OVI\ region was performed, show that
this subtraction is quite helpful, and crucial if \OVI\ is to be
detected at $z\ga 3$ (this correction has not been employed in
previous studies).  Likewise, as shown by the dot-dashed line (where
the weaker doublet component is not used), taking the minumum of the
doublet improves the optical depth recovery considerably.  Additional
noise ($S/N=25$, dotted lines) degrades the correlation at low
$\tau_{\lya}^{\rm rec}$ for $z_{\rm qso}=2.5$, but has little effect
for $z_{\rm qso}=3.5$.  Probably the most dangerous effect is induced by
a significant error in the continuum fit, as this {\em appears} to
generate a false correlation by systematically decreasing \OVI\
optical depths at small $\lya$ optical depths (especially for $z_{\rm
qso}=3.5$ for which continuum fitting errors are large;
see~\ref{sec-optd}).  This, however, is not as problematic as it first
appears, because this effect will only slightly enhance an already-existing
$\tau_{\rm OVI}$ vs.\ $\tau_{\lya}$ trend; we have verified that
it will not create one.

Figure~\ref{fig-odtest_civ} shows the same quantities for \CIV.
Again, \CIV\ is clearly detected at very low density for
$Z=0.01\zsol$; in this case metals are detected at $\tau_{\lya}^{\rm rec} \ga
0.06$ ($\delta \ga 0.4$) for $z_{\rm qso}=3.5$, and at $\tau_{\lya}^{\rm rec}
\ga 0.03$ ($\delta \ga 0.2$) for $z_{\rm qso}=2.5$.  Here, as in
Fig.~\ref{fig-recztest_civ} above, the continuum fitting error becomes
a significant random error source at very small $\tau_{\rm CIV}$ for
$z_{\rm qso}=2.5$, but the recovery is limited by noise for $S/N \la
50$.  The dashed lines show that correction for self-contamination
(also not done in previous studies) is important at low
optical depths.

\begin{figure*}
\vbox{ \centerline{ \epsfig{file=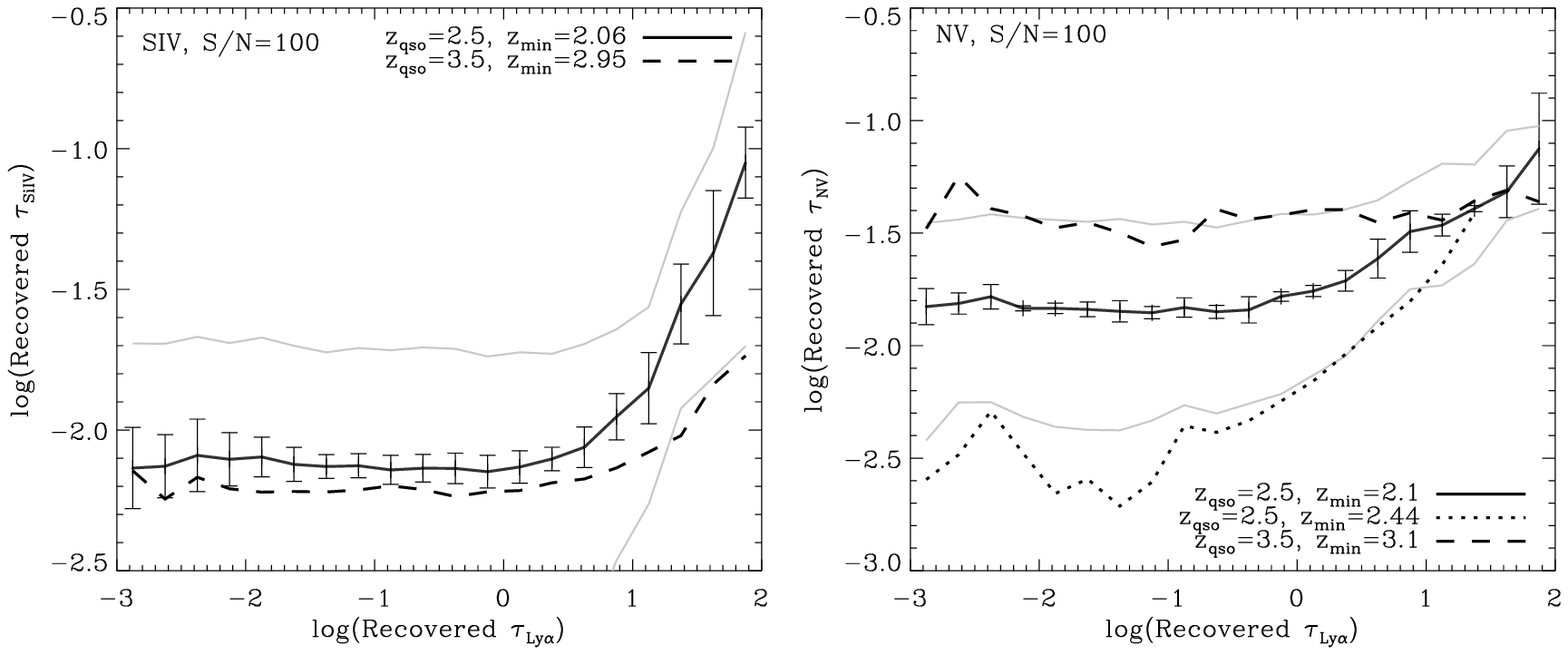,width=18.0truecm}}
\vskip0.15in
\figcaption[]{Recovered SiIV (left panel) or NV (right panel) optical depth
vs.\ recovered $\lya$ optical depth. The metallicity is uniform at
$0.01 Z_\odot$. Solid lines show results for
$z_{\rm qso}=2.5$, with 
$z_{\rm min}=2.06$ (SiIV) or $z_{\rm min}=2.1$ (NV), while dashed lines 
show results for $z_{\rm qso}=3.5$, with $z_{\rm min}=2.95$ (SiIV) or 
$z_{\rm min}=3.1$ (NV).  The dotted line in the NV (right) panel uses
$z_{\rm qso}=2.5, z_{\rm min}=2.44$, which includes only NV pixels redward of
$\lya$. For all curves $z_{\rm max} = z_{\rm qso}$. \footnotesize
\label{fig-odtest_sinv}}}
\vspace*{0.5cm}
\end{figure*}

The results of the POD analysis applied to \SiIV\ and \NV\ are shown
in Figure~\ref{fig-odtest_sinv}.  The left panel shows \SiIV, with the
usual choices of $z_{\rm qso}$, and with $z_{\rm min}$ chosen so that
only \SiIV\ absorption redward of $\lya$ is used.  Despite the lack of
$\lya$ contamination, silicon is detected only at $\tau_{\lya}^{\rm
rec} \ga 3$ (or $\delta\ga 3$).  The recovery is limited primarily by
\CIV\ contamination. Nitrogen V, shown in the right panel, can be
detected well down to $\tau_{\lya}^{\rm rec} \sim 1$ for $z_{\rm
qso}=2.5$ but is nearly undetectable for $z_{\rm qso}=3.5$.  The
dotted line shows the analysis using $z_{\rm min}=2.44$ so that only
the tiny region of \NV\ absorption redward of $\lya$ is used.  The
correlation is strong, but the statistics are poor and this region is
probably too close to the QSO to be of practical use.  Thus, while
useful information can be gleaned from the analysis of \SiIV\ and \NV\
PODs, they are not recoverable at as low $\lya$ optical depths as
\CIV\ and \OVI.

\begin{figure*}
\vbox{ \centerline{ \epsfig{file=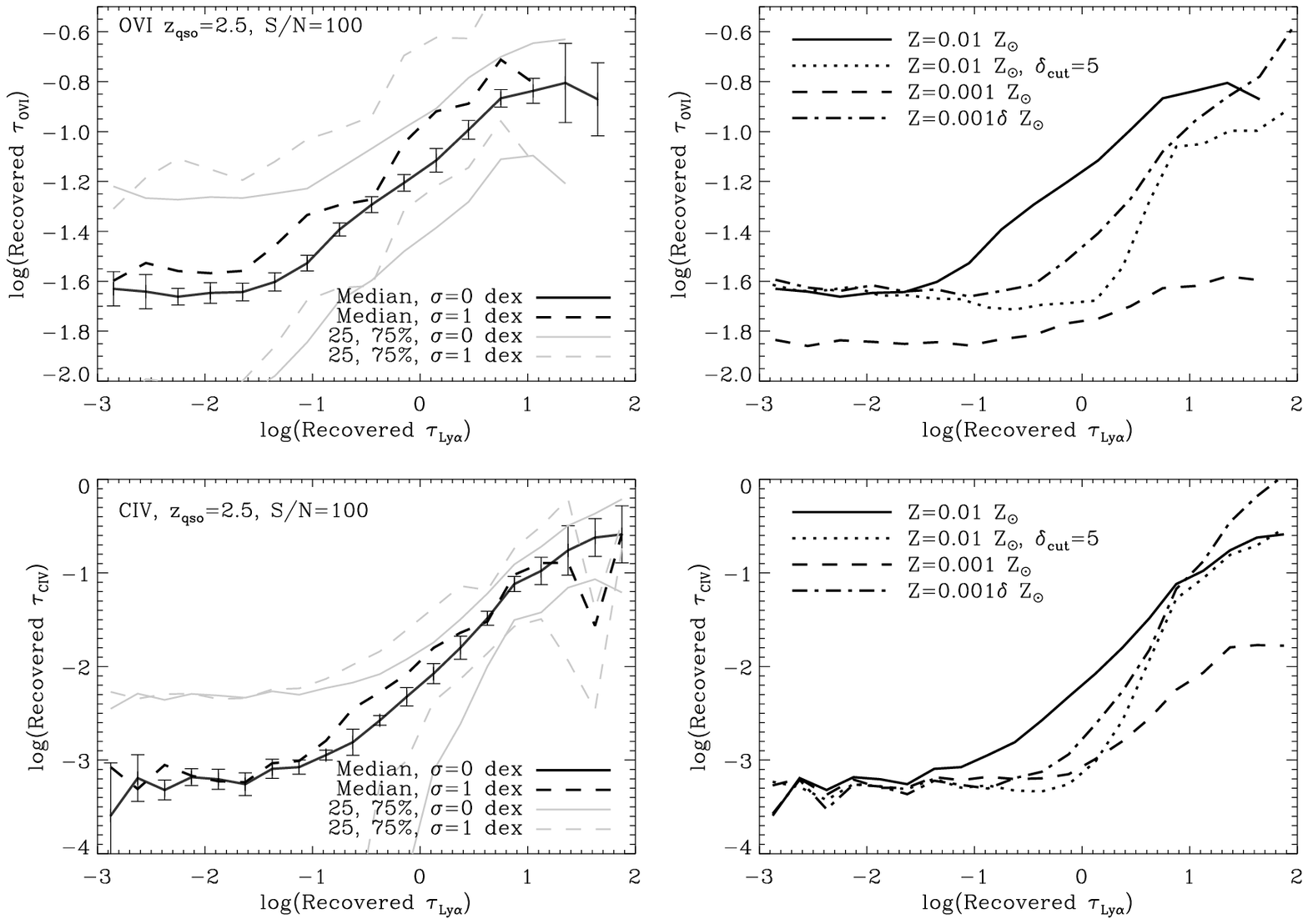,width=18.0truecm}}
\vskip0.15in
\figcaption[]{ \footnotesize Recovered metal optical depth vs.\
recovered $\lya$ optical
depth for OVI (top panels) and CIV (bottom panels) for different metallicity
distributions.  Left panels use a {\em median} metallicity
$Z=0.01\zsol$, but one trial with a uniform metallicity (solid lines),
and one in which there is a 1 dex scatter in $\log(Z)$ on $\sim 1\,$Mpc scales
(dashed lines).  The right panels show trials in which the metallicity
is constant at $Z=0.01\zsol $ (solid) or $Z=0.001\zsol$ (dashed), a trial in
which $Z=const.$ for overdensity $\delta > \delta_{\rm cut} = 5$ and
zero otherwise (dotted), and a trial in which $Z=0.001\delta$ (dot-dashed).
  All panels have $z_{\rm qso}=2.5$, $S/N=100$,
$N_\sigma=3$, $N_{\rm ho}=10$ and $N_{\rm corr}=4$.
\label{fig-cutoff_z2.5}}}
\vspace*{0.5cm}
\end{figure*}

	The overall aim of the POD method is to recover from spectra
the abundances of various metals in low-density gas.
Figure~\ref{fig-cutoff_z2.5} illustrate its effectiveness.  The left
panels re-plot the medians and percentiles for \OVI\ (top) and \CIV\
(bottom) absorption from Figs.~\ref{fig-odtest_ovi}
and~\ref{fig-odtest_civ} as solid lines.  Dashed lines show the
results for a model in which the median metallicity of the gas is
still $0.01\zsol$, but with 1 dex of scatter: the simulation is
divided into cubes $0.6h^{-1}$ comoving Mpc on a side, and each is
assigned a random metallicity drawn from a uniform distribution in $-3
\le \log(Z/\zsol) \le -1$.  Compared to the uniform metallicity case,
the medians are very similar (though slightly higher) and the scatter
slightly larger.  This indicates that the POD method can determine the
overall median metallicity fairly robustly (see also Ellison et
al.\ 2000), but cannot yield strong information about the {\em scatter}
in the metallicity.
	
	The right panels show trials in which the metallicity varies
with gas density.  Solid lines are as in the left panels, and show
results for a uniform metallicity $Z = 0.01\zsol $.  Dotted lines are
trials with $Z=0.01\zsol$ for $\delta > 5$ and $Z = 0 $ for $\delta
\le 5$.  Here the $\tau_Z$ curve flattens, for both metals, at
$\log(\tau_{\lya}^{\rm rec}) \approx 0-0.3$ (or $\delta \approx 4-6$),
confirming that the correlation vanishes at HI optical depths roughly
corresponding to densities at which there are no metals.  The
difference between the solid and dotted lines at $0.2 \la
\log(\tau_{\lya}^{\rm rec}) \ga 0.7$ does, however, indicate that the
presence of metals at $\delta < 5$ affects metal PODs corresponding to
HI PODs with $\log(\tau_{\lya}) > 0$.  The dashed line shows a trial
with $Z=0.001\zsol$.  For carbon the curve is quite similar to that
for $Z=0.01\zsol$, scaled down by $\sim 1$ dex; for \OVI\ the signal
is quite washed out even for larger
$\tau_{\lya}$.  Finally, the dot-dashed line is for metallicity
$Z=0.001\delta\zsol$.  For \CIV\ this curve roughly matches the
$Z=0.01\zsol$ curve at $\log(\tau_{\lya})\approx 0.8$ ($\delta \approx
13$), and matches the $Z=0.001\zsol$ curve for $\log(\tau_{\lya})\sim
-1$ ($\delta \sim 1$), as one would hope (though in the latter case
the values are determined by the noise).  The trend of increasing
metallicity with gas density is apparent in the steep slope of the
dot-dashed line in the \CIV\ plot, but the difference in slope between
this model and the constant 1\% metallicity model with a cutoff is not
very apparent for \OVI.

The results of this section show that the interpretation of the
median pixel optical depth is reasonably straightforward {\em as long
as} noise and other systematic errors are small (as for \CIV) but
becomes somewhat more ambiguous when contamination or
self-contamination are important (as for \OVI).  In the next section we
describe how simulated spectra, which map ``true''
to ``recovered'' optical depths, can be used to directly invert the
pixel optical depths into \CIV/\HI\ and \OVI/\HI\ ratios as a function
of gas density.

\section{Recovering $Z(\delta)$}
\label{sec-invert}

In the preceding sections, we described, tested, and improved the
basic POD recovery method that has been used previously in
analyses of absorption spectra. Despite our improvements, these
recovered PODs are still somewhat affected by contamination, noise, etc.
With our simulated spectra it is, however, possible to extend the basic
method by using the relation -- derived from the simulations --
between true and recovered PODs, to recover a good estimate of the
true metal PODs from observations. Furthermore, if we use the simulations
to determine the relations between density and $\lya$ optical depth,
and between density and temperature, and if we assume a fixed ionizing
background radiation, then we can correct for the
ionization of both the metal and hydrogen and convert the ratio
$\tau_Z/\tau_\lya(\tau_\lya)$ into a metallicity as a function of
density. In this section we describe and
test this procedure.

\subsection{Correcting the optical depth using simulations}

Figures~\ref{fig-recztest_ovi} and~\ref{fig-recztest_civ} show
examples of recovered vs.\ true pixel optical depths in \OVI\ and
\CIV, for particular sets of parameters describing the simulated
spectrum, and for a metallicity $Z(\delta,z)=const=0.01\zsol$ of the
gas as a function of overdensity and redshift. Given the relation
$\tau_Z^{\rm rec}(\tau_Z^{\rm true})$ for a spectrum simulated to have
the same properties as an observed spectrum, the binned recovered
metal PODs (such as those shown in
figs.~\ref{fig-odtest_ovi}-\ref{fig-cutoff_z2.5}) can be ``inverted''
to yield ``true'' PODs, significantly removing the effects of noise,
thermal broadening, contamination by \HI, and continuum fitting
errors.  The remaining effects of self-contamination and contamination
by other metal lines, if important, can also be corrected for {\em
provided} the metallicity used in the simulated spectra (which are in
turn used to perform the inversion) is similar to the metallicity in
the observations.

\begin{figure*}
\vbox{ \centerline{ \epsfig{file=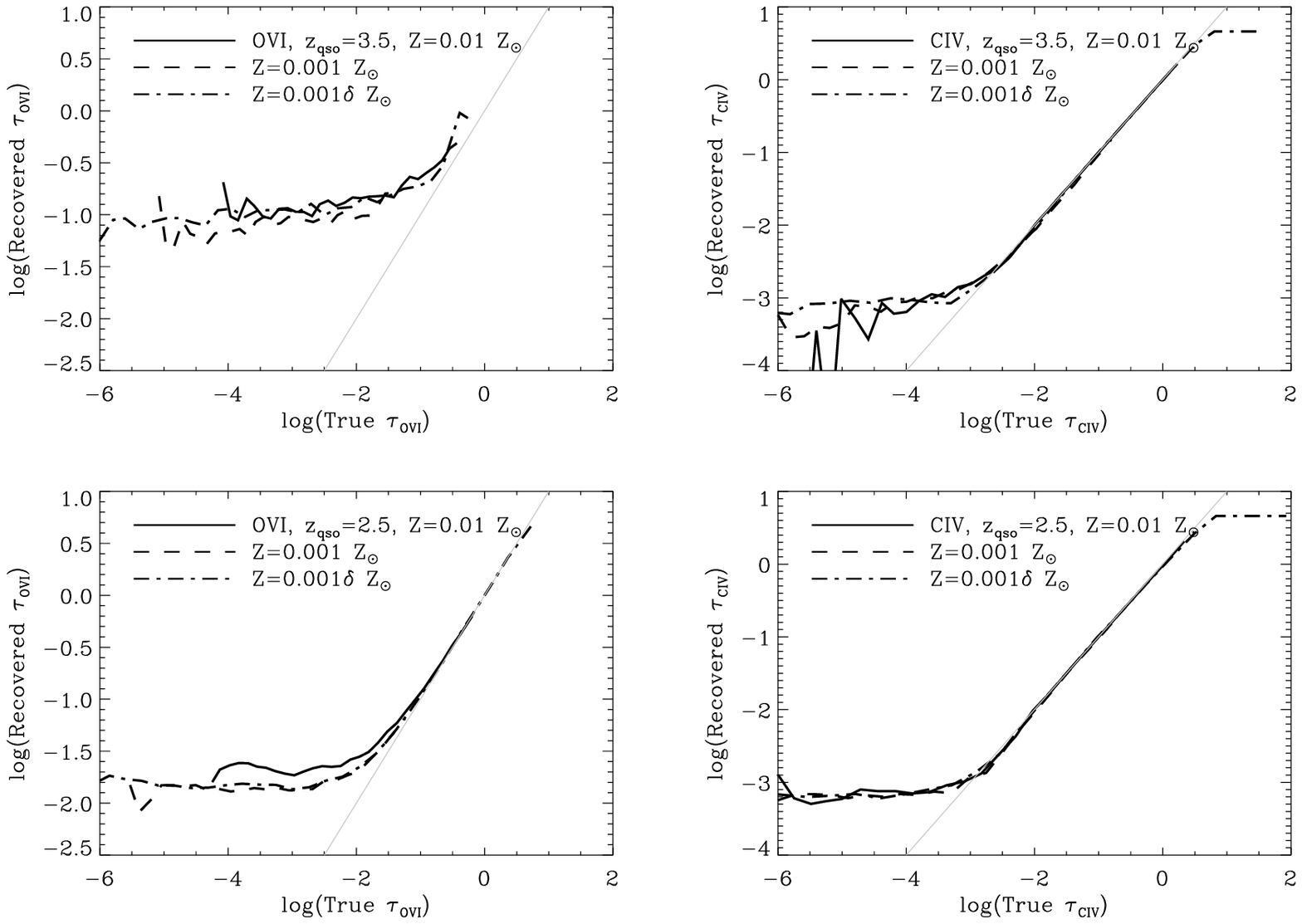,width=18.0truecm}}
\vskip0.1in
\figcaption[]{ \footnotesize Test of the recovery of the pixel optical
depth from the simulated spectra, for different metallicities. Left
panels show OVI, while right panels show CIV.  Top (bottom) panels
have $z_{\rm qso}=3.5(2.5)$.  Recovered vs. true POD is plotted for
uniform metallicities of $10^{-3}\zsol$, $10^{-2}\zsol$, and
$Z=0.001\delta\zsol$. The light solid line indicates perfect recovery.
The flattening at high $\tau_{\rm CIV}$ is due to saturation.
\label{fig-reczztest}}}
\vspace*{0.5cm}
\end{figure*}

Figure~\ref{fig-reczztest} exhibits tests of the importance of metal
line contamination for the recovery of true optical depths for \OVI\
(left panels) and \CIV\ (right panels), for $z_{\rm qso}=3.5$ (top
panels) and $z_{\rm qso}=2.5$ (bottom panels). Curves of $\tau_Z^{\rm
rec}(\tau_Z^{\rm true})$ are plotted for several different choices of
$Z(\delta)$. With the exception of \OVI\ at $z=2.5$, the curves are
virtually identical in their overlap region, indicating that the
deviation from perfect recovery is dominated by factors other than
metal line contamination (\OVI\ at $z=2.5$ has a significant level of
self-contamination; recall that self-contamination was removed for
\CIV). If the curves are independent of metallicity within the range
of $\tau_Z^{\rm rec}$ recovered from the analyzed spectrum, the
inversion can be done in one step; if the assumed metallicity affects
the $\tau_Z^{\rm rec}(\tau_Z^{\rm true})$ obtained, the inversion can
still be done using an iterative approach, as discussed below.

The basic inversion procedure is demonstrated in
Fig.~\ref{fig-inverttest}, which illustrates the process applied to
\CIV\ in a $z_{\rm qso}=2.5$ spectrum with fiducial parameters.  Given
$\tau_{Z,k}^{\rm rec}(\tau_{\lya,k}^{\rm rec})$ in bins $k$ (panel 1)
and the monotonic (hence invertible) curve $\tau_Z^{\rm
rec}(\tau_Z^{\rm true})$ fit though the binned pixels (panel 2), we
are able to compute $\tau_{Z,k}^{\rm inv}(\tau_{\lya,k}^{\rm
rec})\equiv \tau_{Z}^{\rm true}[\tau_{Z,k}^{\rm
rec}(\tau_{\lya,k}^{\rm rec})]$ (hereafter we drop the index $k$ of
the bin). The $\tau_Z^{\rm rec}(\tau_Z^{\rm true})$ relation, which we
compute by averaging over 10 spectra, is indeed monotonic but
increasingly flat at low $\tau_Z^{\rm true}$.  We therefore determine
a value $\tau_{Z,\rm min}^{\rm rec}$ (and $\tau_{Z,\rm min}^{\rm
true}$ ) (shown in light, dotted lines in panels 2 and 3) below which
the curve is effectively flat, and determine an ``error'' $\sigma_{\rm
min}$ on this value by computing the standard deviation of the median
optical depth across the ten realizations for all of the pixels with
$\tau_Z^{\rm true} < \tau_{Z,\rm min}^{\rm true}$. The curve also
varies from one realization of the spectrum to the next; these
variations are quantified in each bin by the standard deviation of the
10 averaged realizations, and two more monotonic functions
$\tau_Z^{\rm rec,\pm}(\tau_Z^{\rm true})$ can be constructed using the
$\pm1\sigma$ values.  The median, $\pm 1\sigma$, and $\pm 2\sigma$
values of $\tau_{Z}^{\rm rec}$ are then inverted into
$\tau_{Z}^{\rm inv}$: the median values are inverted using the mean
$\tau_Z^{\rm rec}(\tau_Z^{\rm true})$, while the $-1\sigma$ and
$-2\sigma$ values are inverted using the $\tau_Z^{\rm
rec,+}(\tau_Z^{\rm true})$ curve and the $+1\sigma$ and $+2\sigma$
values are inverted using the $\tau_Z^{\rm rec,-}(\tau_Z^{\rm true})$
curve. As for the ``flat'' part of the recovered spectrum, we set
$\tau_{Z}^{\rm inv}=0$ when $\tau_{Z}^{\rm rec} < \tau_{Z,{\rm
min}}^{\rm rec}$, and set the $\pm 1\sigma$ and $\pm2\sigma$ values to
zero when they fail to exceed $\tau_{Z,\rm min}^{\rm rec}\mp
\sigma_{\rm min}$.  Projecting the errors this way ensures that: A) we
have included a conservative estimate of the error induced by variance
in the true vs. recovered POD relation from one realization of the
spectrum to the next, and B) the flattening of $\tau_Z^{\rm
rec}(\tau_Z^{\rm true})$ correctly manifests as large negative errors
(effectively upper limits) in $\tau_{Z}^{\rm inv}$ for low values of
$\tau_{\lya}^{\rm rec}$.

The result of this procedure is shown in panel 3.  The plotted points
are our best estimates of the true, uncontaminated $\tau_Z$ versus
$\tau_{\lya}$. These will be reliable to the degree that errors in the
recovered PODs are dominated by noise or \HI\ contamination, which
should be correctly modeled by the simulations.  If the errors are
dominated by continuum fitting errors, the inversion procedure should
correct for this if the same continuum fitting procedure is used (with
equal effectiveness) on the simulated and real spectra. If (as for
\OVI\ at low-$z$) self-contamination is important, the inverted PODs
will be accurate only if the level of metal contamination in the
spectra used in the inversion is similar to that in the real spectrum.

\subsection{Interpreting the optical depth ratios}
\label{sec-intodr}

As mentioned in \S~\ref{sec-interp}, the simulations can be further
used to interpret the PODs in terms of the metallicity $Z(\delta,z)$
of the absorbing gas.  

First, the ratio of the metal to \HI\ optical depth can be converted
into the ratio of the metal ion to \HI\ density, both as a function of
the \HI\ optical depth. For example, for \CIV\ we can compute for each
bin, $$
{n_{CIV} \over n_{HI}} = {\tau_{Z}^{\rm inv} \over  \tau_{\lya}^{\rm
rec}} {(f\lambda)_{HI} \over (f\lambda)_{CIV}}.
$$
Second, the tight relations between gas density and $\lya$ optical
depth, and between temperature $T$ and density, give for each bin in
$\lya$ optical depth, a density and temperature of the corresponding
absorbing gas, denoted respectively by
$\delta_{\lya}(\tau_{\lya}^{\rm rec})$ and
$T_{\lya}(\tau_{\lya}^{\rm rec})$.  These are shown (solid lines)
in panels 4 and 5 for our example inversion.  Given an assumed ionizing
background, these can then be used to compute, for each bin, an
ionization correction $n_H(n_{\rm HI})$ to yield a physical \HI\
density of absorbing gas for each bin. Third and similarly, the
binned, {\em metal} optical-depth-weighted density and temperature,
$\delta_{Z}(\tau_{\lya}^{\rm rec})$ and $T_{Z}(\tau_{\lya}^{\rm
rec})$ (panels 4 and 5, dotted lines) can be used to correct for the
ionization of the metal, to yield an estimate of the density $n_i$ of
the atomic species $i$ ($i$ = C, O, etc.).\footnote{We find that using
$\lya$-weighted 
quantities to correct for H ionization and metal-weighted quantities
to correct for metal ionization works slightly better than using
either set of quantities in both corrections.}  

The ratio $n_i/n_H$ then represents a true metallicity of species $i$
for each bin, yielding (using $\delta_{Z}[\tau_{Z}^{\rm rec}]$) an
estimate of $Z(\delta_Z)$.  The metallicities in our sample inversion
are shown in panel 6, with the dark dashed line representing the true
metallicity.  The good agreement between the measured and true
metallicities demonstrates the effectiveness of the procedure (even
when the metallicity used to derive $\tau_Z^{\rm rec}(\tau_Z^{\rm
true})$ is different from the true metallicity) when the
ionizing background is taken to be the same as used in the spectrum
generation.

The solid and (light) dashed lines indicate the changes in recovered
metallicities induced, respectively, by an overall scaling of $\pm
50\%$ in temperature or density used in the ionization correction;
these are conservative estimates of how much the temperature and
density in the simulations could differ from those of the true IGM.
Such variations induce only small changes in the inferred metallicity
compared with the statistical errors.  

More uncertain is the shape and
spatial variation of the ionizing background, and the importance of
collisional ionization (collisionally ionized gas would have very
different ionization corrections than photoionized gas); here we have
used the model of the UV-background from quasars and galaxies of
Haardt and Madau (2001), and used simulations without feedback from
star formation (i.e., only gravitationally induced shocks are
included) so that the fraction of collisionally ionized gas is
negligable for the redshifts of interest here. The effects of changes
in the assumed ionizing background radiation can be studied by varying
the background used in the inversion, but we will not do this here.

After the recovery of metallicities $Z(\delta)$, the inversion can be
checked or if necessary iterated, as follows.  A new set of simulated
spectra can be generated using the metallicity found in the
inversion,\footnote{In practice this may be done by performing a
power-law or geometric fit to the recovered $Z(\delta)$.  If
self-contamination is important, the inversion is inaccurate if the
metallicity assumed in performing the inversion is very different from
the true metallicity. This effect appears chiefly for $\tau_Z$ below
the dashed line in panel 1 (since, as mentioned in \S~\ref{sec-restest},
this line indicates the level below which the metal contributes
significantly to the absorption). Thus, in the first iteration we fit
$Z(\delta)$ only using 
points with $\tau_Z$ above this dashed line.} and used to generate a new
$\tau_Z^{\rm rec}(\tau_Z^{\rm true})$ (panel 2).  Using this new
relation the inversion can be repeated.  If the results are consistent
with the first iteration, then the recovery was {\em not}
significantly affect by metal-line contamination.  If the first two
iterations differ, the procedure can be repeated until convergence is
obtained.

After convergence, the measured metallicity $Z(\delta)$ can again be
used to generate new simulated spectra. Comparison of the pixel statistics
(e.g., $\tau_Z^{\rm rec}(\tau_\lya^{\rm rec})$) for these spectra and
the original spectra then provides a strong consistency check
on the recovered metallicity, the simulation assumptions, and the
observed spectrum.

\begin{figure*}
\vbox{ \centerline{
\epsfig{file=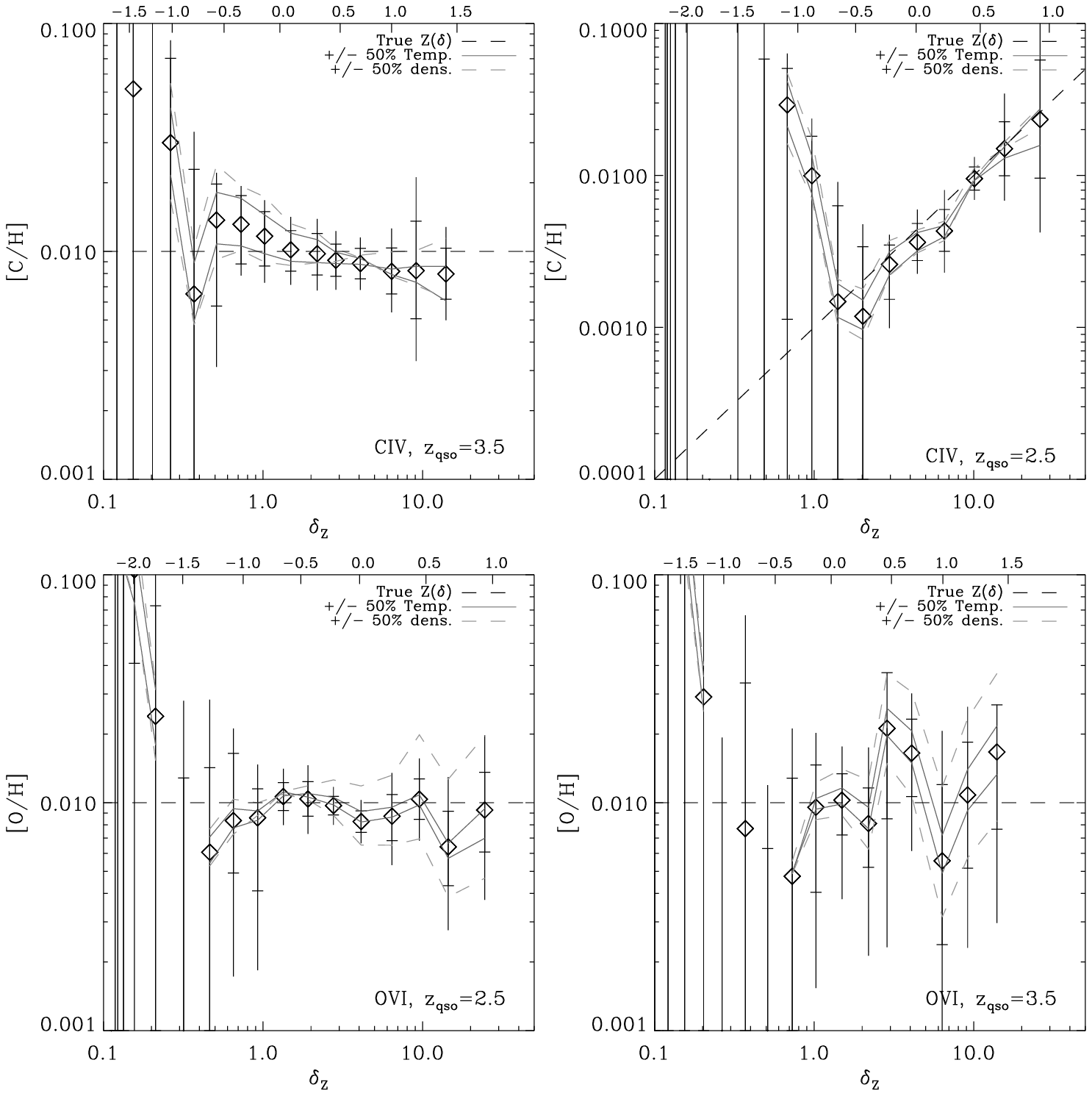,width=18.0truecm}} 
\vskip0.1in
\figcaption[]{
\footnotesize As panel 6 of the previous figure, but for different
elements or metallicities. {\bf Top left:} CIV inversion for $z_{\rm
qso}=3.5$, with $z_{\rm min}=3.3$.  {\bf Top right:} CIV inversion
for $z_{\rm qso}=2.5$, but with $Z=0.001\delta\zsol$ in the analyzed
spectrum, and $Z=0.01\zsol$ in the spectra used for the inversion.
{\bf Bottom left:} OVI inversion for $z_{\rm qso}=2.5$.  {\bf Bottom
right:} OVI inversion for $z_{\rm qso}=3.5$, with $z_{\rm min}=3.3$.
\label{fig-inverttest_multi}}}
\vspace*{0.5cm}
\end{figure*}

The top two panels of Fig.~\ref{fig-inverttest_multi} show the results
of the inversion procedure applied to \CIV\ for $z_{\rm qso}=3.5$
(left panel), and for $z_{\rm qso}=2.5$ with a true metallicity
$Z=0.001\delta\zsol$ (right panel).  The first shows that for $z_{\rm
qso}=3.5$ the inversion is able to constrain the metallicity of even
underdense gas (we find that this holds for $Z \ga 0.001\zsol$); the
second demonstrates that the inversion can cleanly recover a trend in
$Z(\delta)$.  (Note that the $z_{\rm qso}=3.5$ examples use only about
half of the available redshift range; it is worth noting that here --
and in general -- this does not significantly increase the error
bars.)  The bottom two panels show the inversion applied to \OVI, for
$z_{\rm qso}=2.5$ (left panel) and $z_{\rm qso}=3.5$ (right panel).
Despite contamination by \HI, \OVI\ serves as a good tracer of
metallicity for $z_{\rm qso}=2.5$, even in slightly underdense gas.
For $z_{\rm qso}=3.5$ the inversion is not as accurate as for \CIV, but can
still yield a good constraint on the metallicity of overdense gas.

\section{Discussion and Conclusions}
\label{sec-conc}

Studies of the absorption spectra of QSOs at $2 \la z \la 4$ have
revealed that the intergalactic medium is enriched with metals to the
level of $\sim 0.1-1\%$ solar metallicity at overdensities $\delta \ga 5$.
  However, a more accurate
estimate of the metallicity and its spatial variation, as well as the
extent to which it varies with gas density and redshift, are currently
unknown.  This paper is the first in a series that systematically
analyzes a sample of QSO absorption spectra in an effort to glean as much
information as possible about the enrichment of low-density gas at
high redshift.

In the present paper, we have systematically described, tested, and
discussed the ``pixel optical depth" (POD) method of analyzing metal
absorption in QSO spectra.  This method, while still less widely used than
direct line-fitting techniques, is better able to extract information
from low-density regions in which individual metal lines are difficult
or impossible to detect.  The POD technique was developed and
previously employed in QSO absorption studies of \CIV\ (Cowie \&
Songaila 1998; Ellison et al.\ 2000) and \OVI\ (Schaye et al.\ 2000a),
but has suffered from difficulties in the interpretation of the results. Here
we overcome this difficulty by testing the method on spectra generated
from realistic cosmological hydrodynamical simulations that are able
to reproduce the observed statistics of the Ly$\alpha$
forest.\footnote{It should be noted that this is {\em all} that is
required of the simulations, i.e., the results of this paper are not
dependent upon the detailed accuracy of the cosmological simulations,
but only on the simulations' proven ability to roughly match observed
QSO absorption spectra.}  This has allowed us to both assess the
effectiveness of the method, and to refine and calibrate it. The major
improvement to the method developed and tested here is the removal of
higher-order Lyman lines from the \OVI\ region, and the correction of
self-contamination of \CIV\ by its own doublet.  These innovations
significantly improve the accuracy of the \OVI\ and the \CIV\ recovery.  In
the Appendix we gave a compact but complete recipe for
implementing the method.

The most general result of our tests is that the POD technique is very
effective at recovering information about the metallicity of
low-density gas, even where (as for \OVI) the metal lines are severely
contaminated.  More specifically, our tests reveal the following:

\begin{enumerate}
\item{Ly$\alpha$ optical depths of up to several hundred can be
reliably determined using higher order Lyman lines, and these optical
depths in turn give an accurate 
estimate of the density of gas giving rise to the absorption.}
\item{With high-quality spectra (signal-to-noise $\ga 50$) median
\CIV\ optical depths binned in $\lya$ optical depth can be accurately
recovered down to $\tau_{\rm CIV} \sim 10^{-3}$ for both $z_{\rm
qso}=3.5$ and $z_{\rm qso}=2.5$.  Previous analyses were probably
limited by self-contamination; with our correction for this effect, the
recovery is limited by noise and by errors in the continuum fitting.}
\item{The recovery of median \OVI\ PODs is limited primarily by
contamination due to \HI\ lines (and secondarily by continuum fitting
errors). A significant fraction of this contamination can be removed,
allowing a fairly accurate recovery of \OVI\ PODs with $\tau_{\rm OVI}
\ga 10^{-2}$ for $z_{\rm qso}=2.5$, and a useful recovery of median
PODs for $z_{\rm qso}=3.5$.}
\item{The \SiIV\, (limited primarily by \CIV\
contamination) and \NV\, (limited primarily by $\lya$ contamination) PODs
can be usefully recovered for $\lya$ optical depths
($\tau_{\lya} \ga 1)$.}
\item{Using the POD method both \CIV\ and \OVI\ should be detectable
in realistic spectra of $z\sim 3$ QSOs even in gas of the mean cosmic
density, if the metallicity is $\sim 10^{-3}\zsol$, and at even lower
density if the metallicity is higher.}
\item{When applied to simulated spectra with different metallicities
as a function of gas density, the method is able to distinguish models
in which the metallicity is constant from those in which the
metallicity declines significantly with gas density.  The results are
{\em not} very sensitive to spatial variations in metallicity at a
given gas density.}
\item{Most important for accurate recovery of metallicity information
are large wavelength coverage (so that a significant number of higher
order Lyman lines can be used), accurate continuum fitting, and high
signal-to-noise.  However very high S/N is only useful for \CIV\
recovery; obtaining a $S/N \gg 50$ probably does not significantly
improve the amount of information that can be recovered regarding
\OVI.}
\item{Due to different levels of thermal broadening in metals vs.\
hydrogen, the inferred density/temperature of the gas is,
somewhat different when weighted by metal line optical depth than when
weighted by $\lya$ optical depth, especially at relatively high gas
densities.  However, this uncertainty does not appear to significantly
affect the derived metallicities.}
\item{Using spectra generated from cosmological simulations, the POD
method may be extended, as described in~\S~\ref{sec-invert}, to
directly and accurately recover the metallicity of intergalactic gas
vs. its density. The largest uncertainty in this ``inversion'' is the
spectral shape (and possible spatial variations) of the ionizing
background. For a given background, the abundances of C and O can be
measured with errors of at most a factor of a few over at least an
order of magnitude in density, using a single high-quality spectrum.} 

\end{enumerate}

The POD technique, as refined and calibrated in our study should, when
applied to QSO spectra of currently-available quality, yield clear
information about the metallicity of even underdense gas.  This
information will place strong contraints on models of the enrichment
of the IGM, and consequently on feedback and galaxy formation.
 
\acknowledgements

AA and JS are supported in part by a grant from the W.M. Keck
Foundation. TT thanks PPARC for the award of an Advanced
Fellowship. Research was conducted in cooperation with Silicon
Graphics/Cray Research utilizing the Origin 2000 super computer at
DAMTP, Cambridge.

\newpage
\clearpage
\appendix
\section{Summary of the method}
The POD method as we have described and tested above can be
implemented as follows, given an absorption spectrum of a QSO at
redshift $z_{\rm qso}.$
\begin{enumerate}
\item{{\bf Choose redshift range.} Determine $z_\beta \equiv (1+z_{\rm
qso})(\lambda_{\rm Ly\beta}/\lambda_{\rm Ly\alpha})-1$, and choose a
redshift range $z_\beta \le z_{\rm min} < z_{\rm max} \le z_{\rm
qso}$.  For a given transition $i$ with rest wavelength $\lambda_i$,
pixels with wavelength $\lambda_i(1+z_{\rm min}) \le \lambda \le
\lambda_i(1+z_{\rm max})$ can then be analyzed without Ly$\beta$
contamination of the $\lya$ region.}
\item{{\bf Derive Ly$\alpha$ PODs.}  Take
$\tau_{\lya}(z)\equiv-\ln(F)$ where $F(\lambda)$ is the observed normalized flux
at $\lambda=\lambda_{\lya}(1+z)$; mark as ``discarded'' pixels with
$\tau_{\lya} < 0$.

If $F(\lambda) \le 3\sigma_\lambda$ (where $\sigma_\lambda$ is the
normalized noise array), the pixel is saturated. For these pixels, find the
minimum of the 
$\lya$ optical depth corresponding to the optical depth seen in the
available higher-order lines: define
%$$\tau_{\lya}^{\rm rec} \equiv {\rm min}\left\{\tau_{{\rm Ly}n}f_{{\rm
%Ly}\alpha}\lambda_{{\rm Ly}\alpha}/ f_{{\rm Ly}n}\lambda_{{\rm
%Ly}n}\right\},$$
$$\tau_{\lya}^{\rm rec} \equiv {\rm min}\left\{\tau_{{\rm Ly}n}g_{{\rm
Ly}\alpha}/ g_{{\rm Ly}n}\right\},$$ where $g_{{\rm Ly}n} \equiv
f_{{\rm Ly}n}\lambda_{{\rm Ly}n}$, $f_{{\rm Ly}n}$ is the oscillator
strength of the $n$th order Lyman line and $\lambda_{{\rm Ly}n}$ is
its rest wavelength. Compute this, using as many higher-order lines as
fall within the spectrum's coverage, use only pixels that are not ``poorly
detected'', i.e., use those for which $3\sigma_n \le F \le
1-3\sigma_n$ (where $\sigma_n$ is the noise at $\lambda_{{\rm Ly}n}$).
If no higher-order lines are usable, mark the pixel as discarded.}

\item{{\bf Derive metal PODs.} Consider the spectral region covering
all multiplet components of the metal in question for redshift $z_{\rm
min} \le z \le z_{\rm max}$. Then find the optical depth in all
multiplet components, that is, compute $\tau_{Z,k}(z)$ by using the
flux at wavelengths $\lambda_{Z,k} (1+z)$ where $\lambda_{Z,k}$ is the
rest wavelength for the $k$th multiplet component of the metal species
(relevant lines appear in doublets with $k=1,2$).  If the flux $F <
3\sigma_{\lambda_{Z,k}(1+z)}$, the line is effectively saturated; in
that case set $\tau_Z(z)=-\ln(3\sigma_{\lambda_{Z,k}(1+z)})$.

{\bf A. Correct the metal PODs for higher-order Lyman contamination.}
{\em If} the metal lines lie in a region shared with higher-order
Lyman lines (as is the case for \OVI), subtract this contamination
using an interpolation of the $\tau_{\lya}^{\rm rec}$ values derived
in step 2, as follows (otherwise skip to step C):
$$
\tau_{Z,k}(z):=\tau_{Z,k}(z)-\sum_{n=2}{{g_{{\rm
Ly}n} \over g_{{\rm Ly}\alpha}}\tau_{\lya}^{\rm rec}(\lambda\lambda_{\rm Ly\alpha}/\lambda_{{\rm Ly}n})},
$$
where $\lambda=\lambda_{Z,k} (1+z)$, and the sum is over the
 higher-order lines with wavelength covered by the spectrum.  Do this
 for each multiplet component $k$.  Skip now to step C.

{\bf {\em or} B. Correct the metal PODs for contamination and
 self-contamination.}  For lines without Lyman-series contamination, and in
 which self-contamination is dominant (such as \CIV), self-contamination may be
 corrected using the following procedure. 

{\em First correct for contamination by other strong lines:} find all pixels
 with 
$$\exp[-\tau_Z(\lambda)]+3\bar\sigma <
\exp[-(g_2/g_1)\tau(\lambda_s)-(g_1/g_2)\tau(\lambda_d)],$$ where $g_k \equiv
f_{Z,k}\lambda_{Z,k}$,
$\lambda=\lambda_{Z,1}(1+z)$, $\lambda_d=\lambda_{Z,2}(1+z)$,
$\lambda_s=(\lambda_{Z,1}/\lambda_{Z,2})\lambda$, and
$\bar\sigma=\left(
\sigma_{\lambda_s}^2+\sigma_{\lambda_d}^2+\sigma_{\lambda}^2\right)^{1/2}$
is the quadrature sum of the noises at $\lambda_s,\lambda_d$, and
$\lambda$.  Since the optical depth at $\lambda$ should be the sum of
a primary component (with a doublet of strength $g_2/g_1$ at $\lambda_d$)
and a doublet (with a primary component of strength $g_1/g_2$ at
$\lambda_s$), the selected pixels are then very likely contaminated;
discard them so that they are used neither in the analysis nor in the
self-contamination correction. 

{\em Then, correct for self-contamination} using the following algorithm:

1. Begin with the optical depths of the primary component:
$$\tau^{\rm rec}_Z(\lambda)=\tau(\lambda),\ \   \lambda\equiv\lambda_{Z,1}(1+z).$$

2. Subtract the doublets corresponding to these optical depths:
$$\tau^{\rm rec}_Z(\lambda):=\tau(\lambda)-(g_2/g_1)\tau^{\rm
rec}_Z(\lambda\lambda_{Z,1}/\lambda_{Z,2}).$$ (The values of $\tau^{\rm
rec}_Z$ at the specified wavelengths must be interpolated from those
available; for wavelengths lying outside of the available range, take
optical depths from the original full spectrum.)

3. Repeat the second step until convergence occurs (approximately four
more iterations).  The resulting $\tau^{\rm rec}_Z(z)$ is
well-corrected for self-contamination.

{\bf C. Take minimum of doublet} If self-contamination is {\em not}
being corrected (i.e. step B was skipped), take $\tau_Z(z)={\rm
min}\{g_k^{-1}\tau_{Z,k}(z)\}$ using only those components $k>1$ for
which $\tau_{Z,k} > 0$. As above, $g_1=1$, $g_k/g_1 \equiv
f_{Z,k}\lambda_{Z,k}/f_{Z,1}\lambda_{Z,1}$, and $f_{Z,k}$ is the
oscillator strength (for \OVI, \CIV, \NV\ and \SiIV, $g_1=1$,
$g_2=2$).  If $\tau_Z(z) < 0$, set $\tau_Z(z)$ to some fixed value
(e.g., $10^{-5}$) much smaller than the noise.}

\item{{\bf Bin the Ly$\alpha$ and metal PODs.} Create bins
of $-3 \la \log(\tau_{\lya}^{\rm rec}) \la 2$ and compute the median
$\tau_{\lya}$ and $\tau_i$ in each bin.  Errors can be computed by
bootstrap resampling: divide the analyzed region into $5-10$\AA\
chunks (the chunk size should be greater than the line widths), then
compute a large number of realizations in which enough chunks are
randomly selected (each chunk may be selected multiple times) and
concatenated to form a spectrum with a number of pixels equal to the
original. Bin each realization, and take as the error in each bin the
standard deviation from the mean median.}
\end{enumerate}

\end{document}